\documentclass[preprint,12pt,3p]{elsarticle}
\usepackage[utf8]{inputenc} 
\usepackage[T1]{fontenc}    
\usepackage{hyperref}       
\usepackage{url}            
\usepackage{booktabs}       
\usepackage{amsfonts}       
\usepackage{nicefrac}       
\usepackage{microtype}      
\usepackage{lipsum}
\usepackage{amsmath,amssymb,graphicx}
\usepackage{float}
\usepackage{siunitx}
\usepackage{array}
\usepackage{lmodern}
\usepackage{amssymb}
\usepackage{xcolor}
\usepackage{multicol}
\usepackage{multirow}
\usepackage{orcidlink}
\usepackage{amsmath,amssymb,exscale}
\usepackage{amsmath}
\usepackage{amsmath,amssymb,exscale}
\journal{Elsevier}

\begin{document}
\begin{frontmatter}
\title{Elasticity of the PCF-graphene-based single-layer and nanotubes performed by state-of-the-art of reactive classical molecular dynamics simulation method}

\author[Damghan University]{Reza Kalami\corref{author}\,\orcidlink{0000-0003-1194-4210}}
\cortext[author]{Corresponding authors:\ rezakk01@gmail.com  (R. Kalami), saketabi@du.ac.ir (S. A. Ketabi),  josemoreiradesousa$@$ifpi.edu.br (J. M. De Sousa). }

\author[Damghan University]{Seyed Ahmad Ketabi\corref{author}\,\orcidlink{0000-0001-6629-0918}}

\author[IFPI]{José M. De Sousa\corref{author}\,\orcidlink{0000-0002-3941-2382}}

\affiliation[Damghan University]{organization={School of Physics},
            addressline={Damghan University}, 
            city={Damghan},
        country={Iran}}

\affiliation[IFPI]{organization={Instituto Federal de Educa\c c\~ao, Ci\^encia e Tecnologia do Piau\'i -- IFPI},
            addressline={Primavera}, 
            city={São Raimundo Nonato},
            postcode={64770-000}, 
            state={Piauí},
            country={Brazil}}  
                    
\begin{abstract}
A novel two-dimensional carbon allotrope called PCF-graphene has been theoretically proposed. The development of its nanostructured morphology has arrangement a non-aromatic molecule cyclooctatetraene as a precursor (poly-cyclooctatetraene framework). This new carbon allotrope is purely $sp^{2}$-hybridized carbon atoms. In the ﬁrst-principles calculations, PCF-graphene are thermally, mechanically, and dynamically stable. With a ﬁnite thickness of $2.45$\AA, PCF-graphene is a semiconductor with a direct band gap of $0.77$ eV. Exhibits anisotropies in elastic properties, carrier mobility, and optical absorption. Despite having been proposed recently, the study of the mechanical properties of single-layer and nanotubes of PCF-graphene has not been developed yet. Thus, our motivation in this work is to study the mechanical properties of PCF-graphene single-layer and nanotubes. Using state-of-the-art of the fully atomistic classical molecular dynamics simulations method with the use of the interatomic potential ReaxFF implemented in the LAMMPs code, we intend to study the mechanical properties of PCF-graphene $1D$ and $2D$. Our results showed that, the Young's Modulus for  PCF-graphene single-layer for uniaxial strain in the $x$-direction ranges from $5651.7 - 4328.6$ GPa.\AA~ and in the $y$-direction $2408.5 - 1934.4$ GPa\AA~. The Young's modulus of the  PCF-G-NTs ($(0, n)$ and $(n, 0)$) are range $1850.5 - 2603.3$ GPa\AA~ and, $386.25 - 1280.7$ GPa\AA, respectively. The Poisson’s coefficients value are $0.20$ and $0.48$ for PCF-G-NTs $(6,0)$ and $(0,7)$, respectively. We believe that the new results presented to the scientific community in nanoscience can contribute to a theoretical library for future applications of the PCF-graphene nanostructure in the sustained development of new electromechanical devices and new carbon-based materials.
\end{abstract}

\end{frontmatter}

\section{Introduction}
\label{INT}

The carbon allotropy originates several materials, but the only ones occurring in nature are graphite and diamond. Carbon is one of the elements that perform the allotropy phenomenon, that is, it bonds in different ways, forming several simple nanostructures with different physical and chemical shapes and properties. Thus, carbon is able to form many allotropic forms due to its valence \cite{hirsch2010era,falcao2007carbon}.

Due to carbon allotropy, the possibilities of new nanostructures and their applications in nanoscince and naotechnology are possible to be obtained theoretically and experimentally. In recent decades, many more allotropes and forms of carbon have been discovered and researched theoretically and experimentally, including graphene. Graphene, as single sheet of graphite, was synthesized and isolated by mechanical exfoliation \cite{novoselov2004electric}. Graphene is a zero-gap semiconductor, which limits its applications in some electronic devices including digital transistors \cite{withers2010electron}. One of the ways to overcome this problem is to open the electronic gap of graphene, in a controlled way, preserving the electronic properties of graphene \cite{novoselov2004electric,withers2010electron}. Several attempts were made to open the electronic gap of graphene, such as the process of chemical functionalization \cite{elias2009control,eda2010chemically,nair2010fluorinated,flores2009graphene}, strain based gap engineering, applying the electric field and other methods and technologies that can affect the structural symmetries. However, the recently proposals of new allotropic carbon nanostructures which inherently have a certain energy gap with the semiconducting properties and qualities equal to or superior to graphene has opened new perspectives for the design of graphene-based optoelectronic devices. One of such interesting families of carbon allotropes is called Graphyne, represented by 2D-like forming three hybridization states ($sp^{3}$, $sp^{2}$, and $sp^{1}$) hybridized carbon \cite{baughman1987structure,coluci2004ds}. Unlike graphene, the graphyne nanostructurs are semiconductors with direct transitions range from $0.46$ to $1.22$ eV. The elastic modulus of graphyne are range $170 - 240 N m^{-1}$. The bending stiffness of graphyne is estimated to be approximately $1.68$ eV \cite{ivanovskii2013graphynes}. Another recently studied allotrope of carbon is penta-graphene which is composed entirely of carbon pentagons and $sp^{2}$ and $sp^{3}$ hybridized carbon atoms. Sheet of penta-graphene has an indirect band gap of $4.1-4.3$ eV. Due to its morphology configuration (half-thickness being $h = 0.6$\AA), penta-graphene has an elastic properties of the $150.5$ Gpa.nm, UTS of the $38$ Gpa.nm and $ \sigma_{c} = 20$\% \cite{de2021computational}. Theoretical studies developed by classical molecular dynamics method with the interatomic reactive potential ReaxFF, analyzed the effects of temperature on the elastic properties of penta-graphene single-layer. The results obtained showed that penta-graphene thermalized up to $2000$ K, so that the strain rate was reduced to $67$\%, reduction in UTS $35.88 - 11.83$ GPa.nm and Young's Modulus of the $227.15 - 154.76$ GPa.nm \cite{brandao2021atomistic}. The elasticity of penta-graphene-based nanotube has been studied theorically by density functional theory (DFT) and reactive (ReaxFF) classical molecular dynamics method (CMD). The results indicate that the elastic modulus values are in the range of $680 - 800$ GPa, UTS of the $85 - 110$ Gpa and $\sigma_{C} = 18 - 21$\% \cite{de2021mechanical3}.

\begin{figure}[ht!]
\begin{center}
\includegraphics[angle=0,scale=0.48]{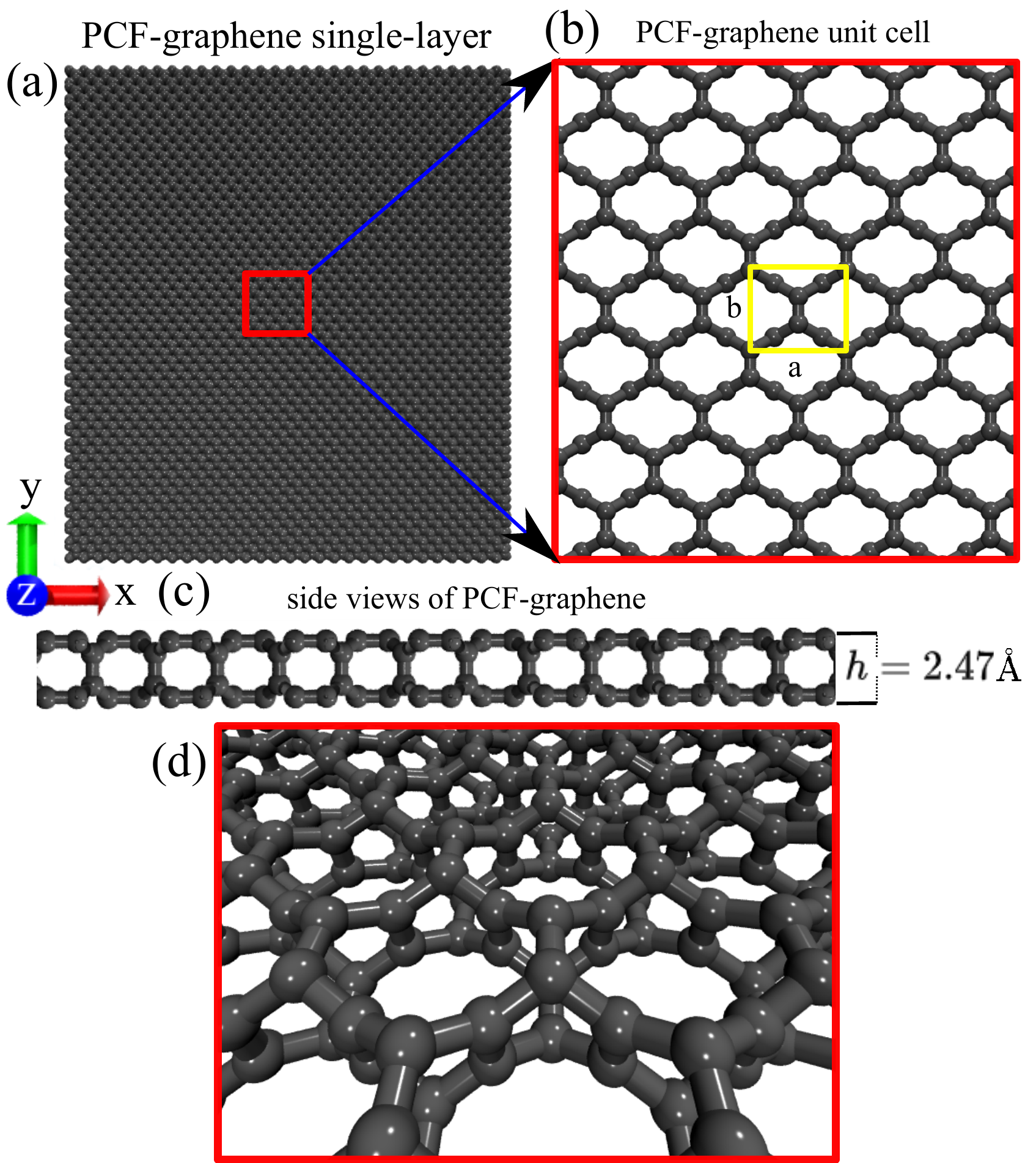}
\caption{\footnotesize{Representative nanostructural fully atomic model of PCF-graphene single-layer  (a) Frontal view in the van der Waals representation. In (b), a zoomed of PCF-graphene single-layer show the carbon atoms and dynamics bonds viewer. The yellow rectangle represents the unit cell, where $a = 4.915$\AA~ and $b = 5.487$\AA. (c) Side view of PCF-graphene single-layer and, (d) a perspective view of PCF-graphene single-layer.}}
\label{FIG:PCF-G:01}
\end{center}
\end{figure}

Another interesting allotrope of carbon is called Poly-Cyclooctatetraene Framework (PCF)-graphene \cite{shen2019pcf}. This new carbon allotrope has a direct band gap of $0.77$ eV. It is similar to the $sp^{2}$-hybridized carbon atoms in graphene (see Fig. \ref{FIG:PCF-G:01}). Calculations performed by first principles ab initio method showed that PCF-graphene is nanostructurally stable (thermally, mechanically, and dynamically), and
exhibits anisotropic optical adsorption and elasticity \cite{shen2019pcf}. Despite having been proposed recently, the study of the elasticity properties of PCF-graphene-based single-layer and nanotubes has not been developed yet. Thus, the purpose of this work is to study the elasticity properties of PCF-graphene single-layer (at temperatures of the $10$ up to $1200$ K) and nanotubes (PCF-G-NTs) at room temperature, performed by state-of-the-art of reactive (ReaxFF) classical molecular dynamics simulation method. The results obtained showed that the Young's Modulus for  PCF-graphene single-layer for uniaxial strain in the $x$-direction ranges from $5651.7 - 4328.6$ GPa.\AA~ and in the $y$-direction $2408.5 - 1934.4$ GPa\AA~, considering the thickness of $3.35$\AA~. The Young's modulus of the  PCF-G-NTs ($(0, n)$ and $(n, 0)$) are range $1850.5 - 2603.3$ GPa\AA~ and, $386.25 - 1280.7$ GPa\AA, respectively. 

In section \ref{INT}, we briefly introduced the preliminary concepts and the motivations for the development of this theoretical work. In sections \ref{CM} and \ref{CM2} we present the computational methodology used to model the nanostructures of  PCF-graphene-based single-layer and nanotubes. The results and discussion are presented in Section \ref{results}  followed by conclusions and remarks in Section \ref{conclusion}.

\section{Computational Methodology}
\label{CM}

The reactive molecular dynamics simulations method were carried out to study the tensile stress/strain behavior of PCF-graphene single-layer and nanotube (PCF-G-NTs). These reactive molecular dynamics simulations were performed using the LAMMPS (Large-scale Atomic/Molecular Massively Parallel Simulator) \cite{plimpton1995fast} code with the interatomic reactive Force Field (ReaxFF) \cite{mueller2010development}. ReaxFF is a interatomic reactive force field developed by van Duin, Goddard III and co-workers for using in molecular dynamics method calculations. 

The system energy is divided into partial energy contributions, such as, bond distance $(E_{bond})$, the over-coordination $(E_{over})$, the under-coordination $(E_{under})$, the valence $(E_{val})$, the penalty for handling atoms with two double bonds $(E_{pen})$, the torsion $(E_{tor})$, the conjugated bond energies $(E_{conj})$,  the van der Waals $(E_{vdW})$, and coulomb interactions $(E_{co})$. The ReaxFF can handle bond formation and dissociation, making and breakingbonds, as a function of bond order values. ReaxFF was parametrized whith DFT calculations, being the average deviation between the heats of formation predicted by the theory and by the experiment equal to $2.8$ and
$2.9$ $kcal$ $mol^{-1}$, for non-conjugated and conjugated systems. ReaxFF has been successfully used in the study of elastic properties of nanostructured systems \cite{de2016mechanical,de2019elastic,de2016torsional}, among other nanostructures. 

The nanostructural model of the PCF-graphene single-layer (shown in Figure \ref{FIG:PCF-G:01}) consists of $14400$ carbon atoms and has dimensions of $146.22 \times 163.23$\AA$^{2}$. Nanostructural details of the (PCF-G-NTs) are presented in Table \ref{tab:PCF-NTs-tab}. Before coupling the thermostat chain on the PCF-graphene single-layer and nanotubes (PCF-G-NTs) for thermodynamic equilibrium before starting the mechanical stretching process, all nanostructures studied in this work are initially subjected to carbon atoms restrained energy minimization with respect to both the torsion angles and the atomic Cartesian co-ordinates. Before starting the stretch processes, in order to eliminate any residual stress present on the all nanostructures, we perform constant $NPT$ integration thermalization carried at null pressure using a Nose/Hoover pressure barostat during $10000$ fs \cite{evans1985nose}. Then, the tensile mechanical were performed by stretching the PCF-graphene single-layer and nanotubes (PCF-G-NTs) until nanofracture within NVT ensemble (Nose/Hoover temperature thermostat) at room temperature during $10000$ fs. In all reactive molecular dynamics simulations we use a timestep of $0.05$ fs. We used a constant engineering tensile strain rate $\delta = 10^{-6}fs^{-1}$. 

The elastic properties of the PCF-graphene single-layer and nanotubes (PCF-G-NTs) were analyzed by the stress-strain relationship, where the engineering strain, $\varepsilon$, under tension is defined as
\begin{eqnarray}
\label{epis}
\centering
\varepsilon = \frac{\zeta - \zeta _{0}}{\zeta _{0}}= \frac{\Delta \zeta }{\zeta _{0}},
\end{eqnarray}
where $\zeta_0$ and $\zeta$ are the length of the structure before and after the dynamics of deformation, respectively. The {\it per-atom} stress tensor of each carbon atom are calculated by~\cite{muller2011fundamentals}:
\begin{eqnarray}
  \sigma_{\alpha \beta }={\Gamma }^{-1}\sum_{i}^{N}\left (m_{i}v_{\alpha i}v_{i\beta } + r_{i\alpha }f_{i\beta } \right ) , 
\end{eqnarray}
where $\Gamma$ is the atom volume, $N$ the number of atoms, $m_i$ the mass of carbon atoms, $v$ the velocity, $r$ the coordinates of the carbon atoms and $f_{i\beta}$ is the $\beta$ component of the force acting on the $i$-{\it th} atom. In order to perform a more detailed analysis of the distribution of
stress along the structure during the fracture process, we also calculated the quantity known as {\it von Mises stress}, $\sigma_{vM}$, which is mathematically given by~\cite{muller2011fundamentals}:
\begin{eqnarray}
  \sigma_{vM}=\sqrt{\left [ \frac{(\sigma_{xx}-\sigma_{yy})^{2}+
      (\sigma_{yy}-\sigma_{zz})^{2}+(\sigma_{zz}-\sigma_{xx})^{2}+
      6(\tau_{xy}^{2}+\tau_{yz}^{2}+\tau_{zx}^{2})^{}}{2} \right ]} , 
\end{eqnarray}
where $\sigma_{xy}$, $\sigma_{yz}$ and $\sigma_{zx}$ are shear stress components. The $\tau_{xy}$, $\tau_{yz}$ and $\tau_{zx}$ are shear stress components. It allows a dynamical visualization (qualitatively) of where the stress is accumulated and dissipated during the stretching and nanofracture process along the whole nanostructure of PCF-graphene-based single-layer and nanotubes (PCF-G-NTs).

\subsection{Calculation of Poisson's ratio of the PCF-graphene-based nanotubes (PCF-G-NTs)}

The cylindrical nanostructural shape of PCF-graphene-based carbon nanotubes (PCF-G-NTs) theoretically investigated in this research work, the PCF-G-NTs are subjected at uniaxial strain load in $z$-direction, ($\epsilon $), without the radial constraint. Thus, the evaluated Poisson's ratio is mathematically given by the following expression \cite{wang2005size}:

\begin{eqnarray}
    \nu = - lim _{\varepsilon = 0} \left( \frac{\varepsilon^{'}}{\varepsilon} \right),
\end{eqnarray}
where, $\nu$ Poisson's ratio, $\varepsilon^{'}$ radial strain and $\varepsilon$ uniaxial strain applied in PCF-G-NTs. The Poisson's ratio, $\nu$, of PCF-G-NT $(6,0)$ and $(0,7)$ are calculated by \cite{wang2005size}:

\begin{eqnarray}
    \nu = - \frac{\varepsilon _{R}}{\varepsilon _{z}},
\end{eqnarray}
The radial strain ($\varepsilon _{R}$) and uniaxial strain ($\varepsilon _{z}$) are defined in terms of strain mechanical load. The uniaxial strain in $z$-direction can be measured from the rate of deformation in PCF-G-NTs $(6,0)$ and $(0,7)$. The length ($L$) of PCF-G-NTs $(6,0)$ and $(0,7)$ is divided into equal-size slabs with thickness ($\Delta L = \frac{L}{n}$). The uniaxial strain load in $z$-direction is then calculated by:
\begin{eqnarray}
\epsilon_{R} = \frac{R^{(average)}_{\varepsilon_{z}} - R^{(average)}_{0}}{R^{(average)}_{0}},
\end{eqnarray}
where, $R^{(average)}_{\varepsilon_{z}}$ is the average PCF-G-NTs $(6,0)$ and $(0,7)$ radius at the uniaxial strain load $\varepsilon_{z}$. The $R^{(average)}_{0}$ is the average of the PCF-G-NTs radius at the equilibrium, $\varepsilon_{z}=0$, where:
\begin{eqnarray}
R^{(average)}_{\varepsilon_{z}} = n^{-1}\sum_{i=1}^{n} R^{(slab)}_{i}|_{\varepsilon_{z}}.
\label{Eq:07}
\end{eqnarray}

In Eq. \ref{Eq:07}, $R^{(slab)}_{i}|_{\varepsilon_{z}}$ is the average radius of the {\textit{i}}-esim circular slab of the PCF-G-NTs $(6,0)$ and $(0,7)$ at uniaxial strain $\varepsilon_{z}$. $R^{(slab)}_{i}$ is the average of the distance of each slab composed by carbon atoms to its center of mass. Is calculated by:
\begin{eqnarray}
R^{(slab)}_{i} = M^{-1}\sum_{\alpha 
= 1}^{N} r_{\alpha},
\label{Eq:08}
\end{eqnarray}
where, $r_{\zeta}$ is given by the mathematical expression:
\begin{eqnarray}
r_{\alpha} = \sqrt{\left( x_{\beta} - x_{CM}  \right)^{2} + \left( y_{\beta} - y_{CM}  \right)^{2}}.
\label{Eq:09}
\end{eqnarray}
The Eq. \ref{Eq:09}, $x_{\beta}$ and $y_{\beta}$ are the planar coordinates of the $\alpha$-esim carbon atoms, where $M$ is defined by carbon atoms per slab. The coordinates $x_{CM}$ and $y_{CM}$ are the mass center of each slab of PCF-G-NTs $(6,0)$ and $(0,7)$. We consider $\Delta L = 1.5$\AA~. There is reasonable minimum number of atoms at each slab of the radius of the PCF-G-NTs $(6,0)$ and $(0,7)$ \cite{brandao2023first}.

\section{Generation of PCF-graphene-based nanotubes - PCF-G-NTs}
\label{CM2}
The computational construction of PCF-G-NTs is initially done by defining the chiral vector (see Fig.\ref{Fig:PCF-G-NT}):
\begin{equation}
    \textbf{C}_h=n\textbf{a}+m\textbf{b},
    \label{eq:chiral-vec}
\end{equation}
where $\textbf{a} = 4.915$\AA~ and $\textbf{b} = 5.487$\AA   are the lattice vectors and $n,m$ are integers that characterize the type of chirality. Through this vector we can obtain the diameter of the nanotube:
\begin{equation}
    d_t=\dfrac{|\textbf{C}_h|}{\pi}.
\end{equation}
The translational vector is perpendicular to the chiral vector, given by (see Fig.\ref{Fig:PCF-G-NT}):
\begin{equation}
    \textbf{T}=t_1\textbf{a}+t_2\textbf{b},
\end{equation}
where $t_1,t_2$ are integers that can be obtained using the inner product $\textbf{C}_h\cdot\textbf{T}=0$, thus
\begin{equation}
    \dfrac{t_1}{t_2}=-\dfrac{m}{n},
\end{equation}
knowing that $\textbf{a}\cdot\textbf{b}=0$ for the rectangular lattice. The length of the nanotube is given by $L=|$\textbf{T}$|$. The $\textbf{C}_h$ (see Fig.\ref{Fig:PCF-G-NT}) and $\textbf{T}$ vectors therefore delimit the PCF-G-NTs unit cell. The PCF-G-NTs rectangular lattice allows the construction of PCF-G-NTs zigzag-like, $(n,0)$ or $(0,n)$ chirality. The structural parameters of PCF-G-NTs studied in this work can be viewed in Table \ref{tab:PCF-NTs-tab}.

\section{Results and discussions}
\label{results}
\begin{figure}[ht!]
\begin{center}
\includegraphics[angle=0,scale=0.35]{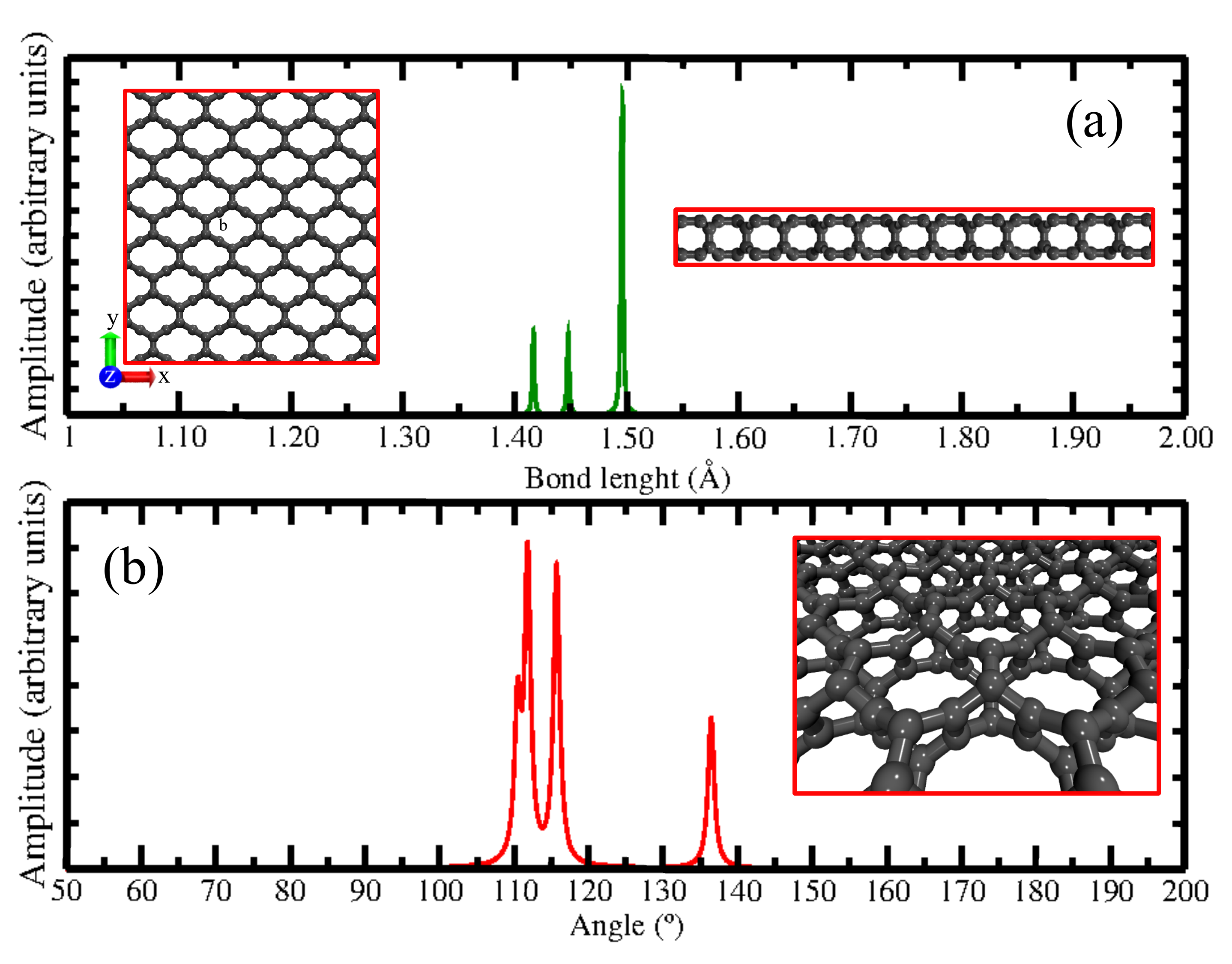}
\caption{\footnotesize{Graphical representation of the evolution of bond length (a) and bond angle (b) of the nanostructural PCF-graphene minimized by ReaxFF set parameter \cite{mueller2010development}.}}
\label{FIG:PCF-G-Lenght_Bonds_Angles}
\end{center}
\end{figure}

\begin{figure}[hp!]
\begin{center}
\includegraphics[angle=0,scale=0.25]{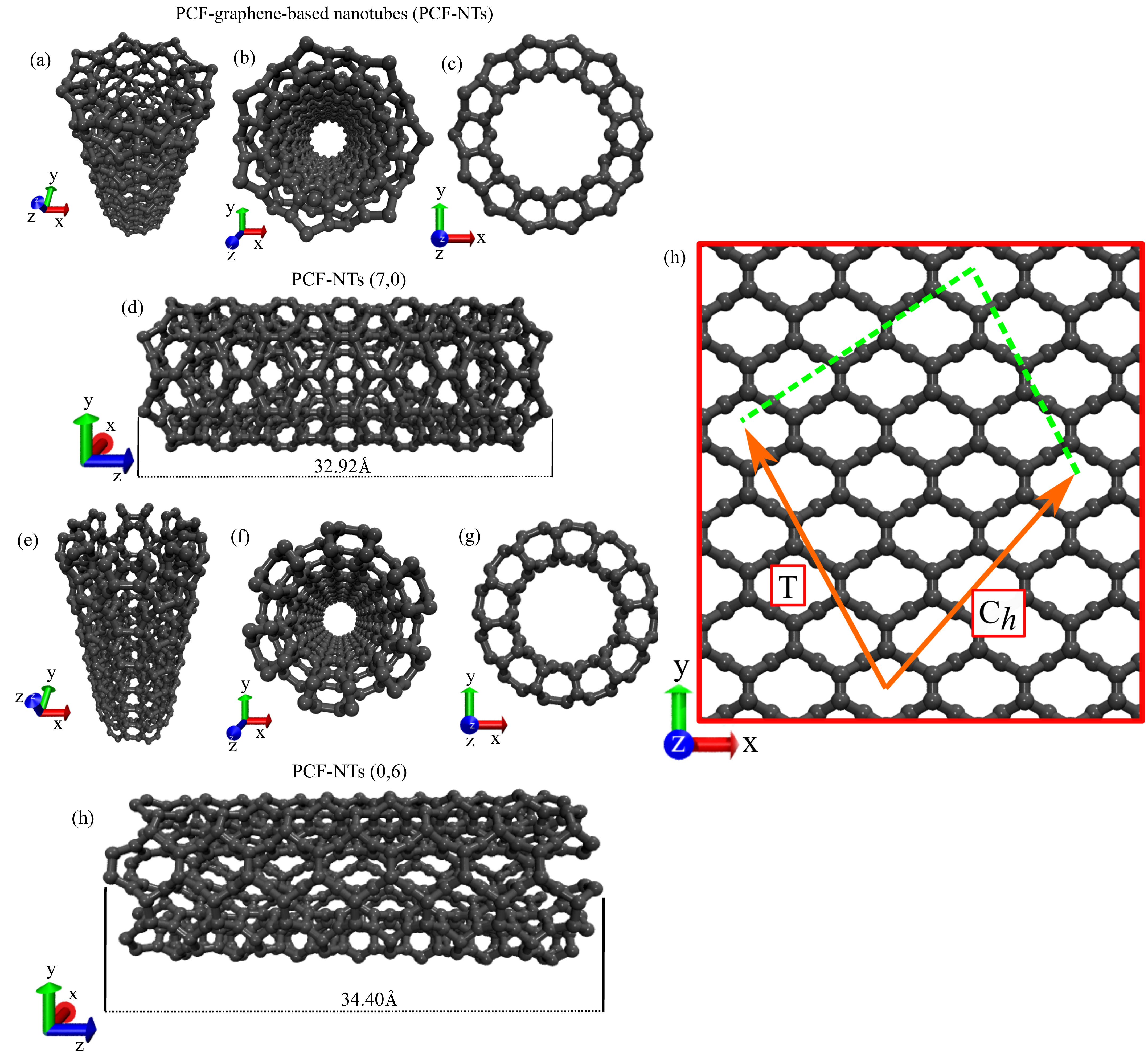}
\caption{\footnotesize{Fully atomistic model illustration of the PCF-graphene nanostructure based nanotubes PCF-G-NT. In viewer module rendered by VMD of the PCF-G-NT $(0,7)$ in (a) longitudinal, (b) perspective, (c) orthographic and (d) length, respectively. In viewer module rendered by VMD of the PCF-G-NT $(6,0)$ in (a) longitudinal, (b) perspective, (c) orthographic and (d) length, respectively. In (h) representation of the chiral ($\mathbf{C}_h$) and translational ($\mathbf{T}$) vectors for a PCF-G-NTs.}}
\label{Fig:PCF-G-NT}
\end{center}
\end{figure}

Here, we represent the results of reactive (ReaxFF) classical molecular dynamics (CMD) simulations of the stretching dynamics of PCF-graphene single-layer and PCF-graphene-based nanotubes (PCF-G-NTs) in order to investigate the elastic properties and fracture pattern of these nanostructures in $1D$ and $2D$ morphology dimensions. In Fig. \ref{FIG:PCF-G-Lenght_Bonds_Angles}, we present the bond length distribution analysis for PCF-graphene nanostructure configurations. Notably, the bond lengths exhibit a distribution spanning
the range of $1.42-1.50$\AA~. The angle distribution analysis for PCF-graphene nanostructure exhibit a distribution spanning the range of $112-136$. The Young's modulus, ultimate tensile streght (UTS) and critical strain values obtained by CMD for PCF-graphene single-layer and PCF-G-NTs studied in the research work are presented in Table \ref{tab:elastic-val}.

Complete nanostructural failure of PCF-graphene single-layer ($146.22 \times 163.23$)\AA$^{2}$ ($14400$ carbon atoms) ranges from $11.12 - 14.87$\% ($x$-direction) and $16.51 - 33.37$\%~ ($y$-direction) of strain. The PCF-G-NTs $(n,0)$ ranges from $22.22 - 39.27$\% and $(0,n )$ $12.23 - 25.87$ \% of strain, respectively. On the other hand, different critical strains were observed when comparing graphene single-layer and conventional CNTs. Where the critical strain of graphene single-layer (($90 \times 90$)\AA$^{2}$ whith $3656$ carbon atoms) with uniaxial stretching applied in the $x$ direction is $10.96$\% and in the $y$ direction is $13.24$\%. For conventional nanotubes these values change from $16 - 18$\%, results obtained with the same computational methodology developed here in this work \cite{lee2008measurement,de2021nanostructures}. We attribute the differences due to the nanostructural morphology between PCF-graphene and graphene single-layer, where the PCF-graphene single-layer presents in its nanostructure all carbon atoms threefold coordinated (see Fig. \ref{FIG:PCF-G:01}). The graphene single-layer, on the other hand, presents densely packed hexagonal carbon atoms in its nanostructural morphology. Thus, like conventional CNTs formed by rolled up a single layer of graphene, also presents a densely packed nanostructural morphology \cite{lee2008measurement,de2021nanostructures}. Therefore, the nanostructural morphology between PCF-graphene and graphene single-layer and nanotubes, present significant differences in their elastic deformation and significantly affect
its mechanical properties. The Young's modulus between single-layer PCF-graphene and nanotubes also show significant differences. The results obtained in our computer simulations show that the Young's Modulus for single-layer PCF-graphene for uniaxial strain in the $x$ direction ranges from $5651.7 - 4328.6$ GPa.\AA~ and in the $y$ direction $2408.5 - 1934.4$ GPa.\AA~, for temperatures ranges of the $10$, $300$, $600$, $900$ and $1200$ K. Comparing with the Young's Modulus of graphene single-layer the value is $3350$ GPa.\AA~, considering the thickness of $3.35$\AA \cite{lee2008measurement}. We can observe that for deformation in the x direction, the Young's Modulus for the single-layer PCF-Graphene is $40.99$\%~ larger than the graphene single-layer. For deformation in the $y$ direction Young's modulus is $27.78$\% smaller than the graphene membrane. Our results show that Young's modulus of PCF-graphene single-layer presents changes in its values due to the effects of temperature. In the $X$ direction the values decrease in $23.41$\%~ while in the $Y$ direction the values decrease in $19.68$\%~. Lu, P. L. (2008), investigated using an empirical force-constant model, the mechanical properties of single wall conventional carbon nanotubes. The results showed that the Young's Modulus CNTs is $970$ GPa (or $3249.5$ GPa.\AA~, whith thickeness of $3.35$\AA) \cite{lu1997elastic}. Hernandez, E. \textit{et. al} (1998), using a non-orthogonal tight-binding formalism, obtained in research results the Young's Modulus value of CNTs are of the $1080$ Gpa (or $3618$ Gpa.\AA~, whith thickeness of $3.35$\AA). \cite{hernandez1998elastic} De Sousa (2021), using the reactive (ReaxFF) classical molecular dynamics method, obtained the Young's Modulus of CNT of the $955$ GPa (or $3199.25$ Gpa.\AA~, whith thickeness of $3.35$\AA) \cite{de2021nanostructures}. Comparing these values, the PCF-G-NTs present a Young's Modulus around $18.63$\% smaller than the conventional CNTs. The utimate tensile streght of PCF-graphene single-layer and PCF-GNTs ($(0,n)$ and $(n,0)$) are range $1934.4 - 5651.7$ GPa\AA~, $1850.5 - 2603.3$ GPa\AA~ and, $386.25 - 1280.7$ GPa\AA, respectively. All values of the mechanical properties of PCF-graphene-based nanotubes are presented in Table \ref{tab:elastic-val}.

The mechanical properties PCF-graphene single-layer (unaxial strain in $x$-direction and $y$-direction at room temperature) and PCF-GNTs ($(0,n)$ and $(n,0)$) at room temperature)  are presented follow. The nanostructural failure (fracture pattern) processes can be better understood following the fully atomistic molecular dynamics of the temporal evolution of the von Mises stress distributions from the snapshots of the tensile stretch, see Figs. \ref{FIG:PCF-G-X:02}, \ref{FIG:PCF-G-Y:03} and \ref{FIG:PCF-GNTs_snapshots}, respectively. From the results of the computational simulations presented in these figures, it is possible to observe a high stress accumulation (in red color) along the $1D$ and $2D$ PCF-graphene nanostructures from applied stress through uniaxial stretching of the computer simulation box. This process computational change the shape of the simulation box during a reactive (ReaxFF) molecular dynamics simulations run by \textit{constant engineering strain rate (see section \ref{CM}: computational methodology)}. The chemical bonding of PCF-graphene resembles that in the product of a cycloaddition reaction. The space group of the optimized lattice is $Cmmm$ (space group number of  65), and the carbon atoms occupy two nonequivalent Wyckoﬀ positions \cite{shen2019pcf}.

\begin{figure}[ht!]
\begin{center}
\includegraphics[angle=0,scale=0.20]{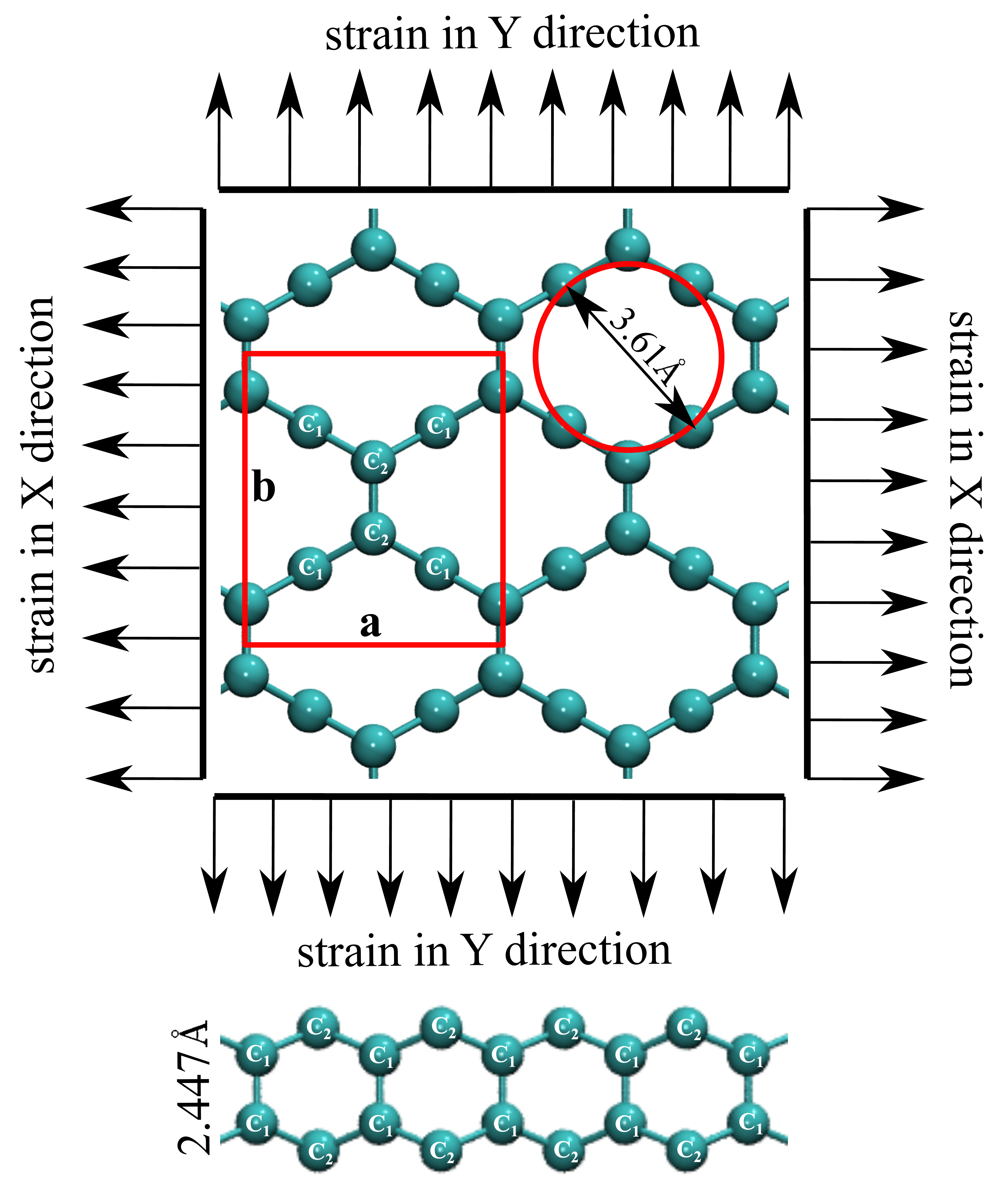}
\caption{\footnotesize{Schematic representation of strain applied directions $X$ and $Y$ of the PCF-graphene single-layer. A top and side views of the atomic conﬁguration of PCF-graphene single-layer. The full line rectangle in red color represents the unit cell (\textbf{a}=$4.915$\AA~ and \textbf{b}=  $5.487$\AA). The lenght bonds $C_{1} - C_{1}$ is $1.42$\AA~, $C_{2} - C_{2}$ is $1.45$\AA~,  $C_{2} - C_{1}$ is $1.50$\AA~ and  $C_{1} - C_{2}$ is $1.50$\AA~.}}
\label{FIG:PCF-G-BONDS-STRAIN}
\end{center}
\end{figure}

\begin{figure}[ht!]
\begin{center}
\includegraphics[angle=0,scale=0.34]{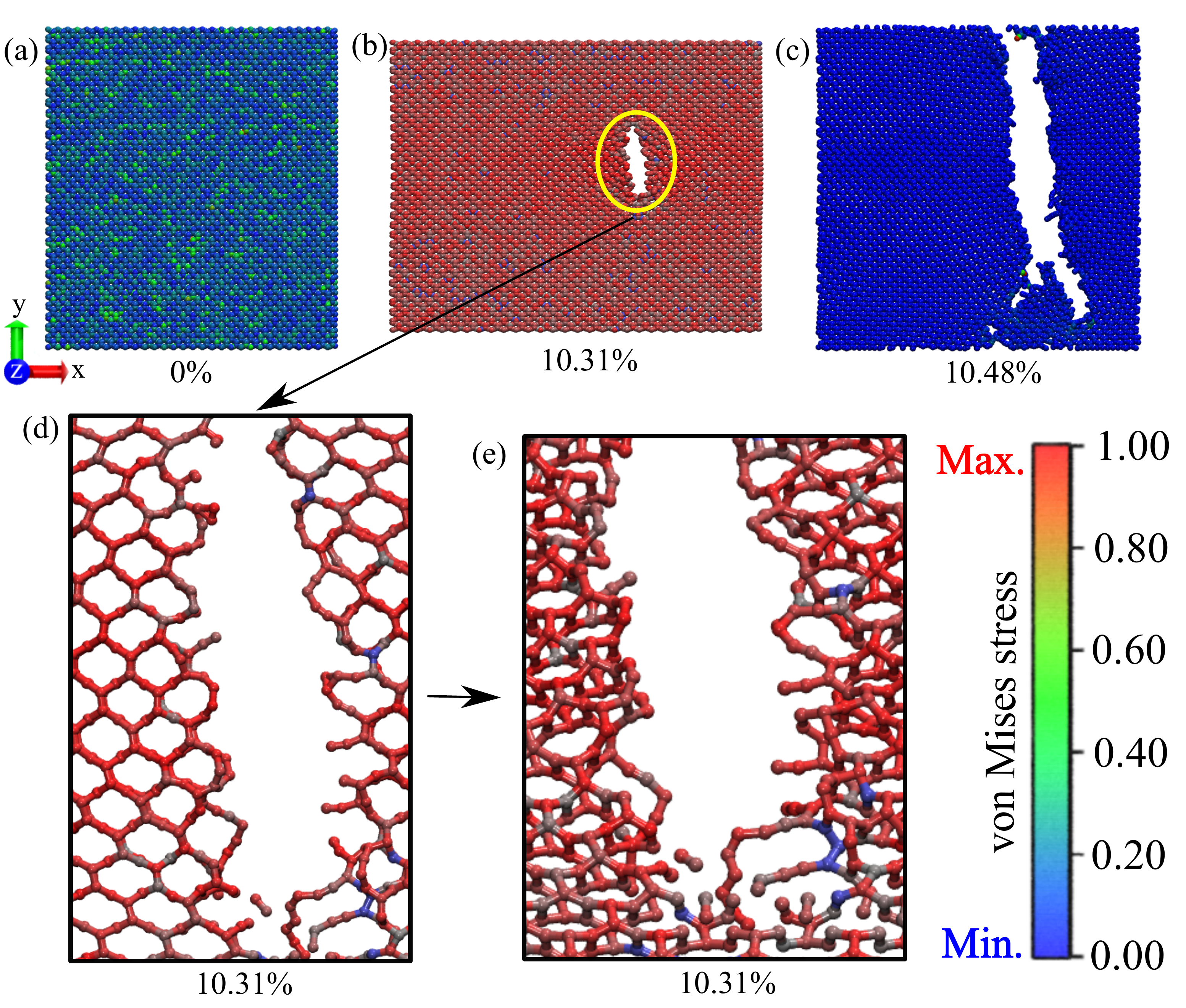}
\caption{\footnotesize{Representative nanostructural fully atomic model of PCF-graphene single-layer under uniaxial stress load in $X$ direction at room temperature. In (a) at $0$\% of strain, in (b) initiation of nanofracture with the breaking of some chemical bonds at $10.31$\% of strain, in (c) the single-layer completely nanofractured in the morphological configuration divided into two parts at $10.48$\% of strain, in (d) a zoomed view showed the break of some chemical bonds $C - C $ at $10.31$\% of strain and (e) a zoomed view in perspective. In the sidebar located on the right side, the red color indicates high-stress accumulation, while blue color indicates low-stress accumulation in the PCF-graphene single-layer.}}
\label{FIG:PCF-G-X:02}
\end{center}
\end{figure}

In Fig.\ref{FIG:PCF-G-X:02}, we present the result of the computational simulation of PCF-graphene single-layer under uniaxial stretching deformation in the $x$ direction at room temperature. Thus, snapshots of the computational simulations were generated before and during the stretching until the complete mechanical fracture physically characterized by the division of the PCF-graphene single-layer ($x$-direction) into two parts. Analyzing the fracture pattern, we can observe in the results obtained in the computational simulations that the broken of the chemical bonds $C -C$ are in the bonds of type $C_{1} - C_{2}$ and $C_{2} - C_{1}$  approximately parallel to the direction of the uniaxial deformation of the PCF-graphenbe single-layer (see Fig.\ref{FIG:PCF-G-BONDS-STRAIN}). In all reactive molecular dynamics simulations results, the break bonds of the $C_{2} - C_{2}$ and $C_{1} - C_{1}$ was not identified in computational simulations with strain applied in $x$ direction. In Fig.\ref{FIG:PCF-G-X:Diameter}, we can observe the pore evolution of PCF-graphene single-layer under uniaxal strain load applied in $x$-direction. We can see that porous large deformations but after nanofracture of membrane ($3.61$\AA~ up to $3.90$\AA~), return to original nanostructural shape, in approximately $3.58$\AA (see Fig.\ref{FIG:PCF-G-X:Diameter}). As Fig.\ref{FIG:PCF-G-Y:03} shows the representative nanostructural fully atomic model of PCF-graphene single-layer snapshots under uniaxial stress load in $y$-direction at room temperature. In plots (a) - (c), the stretching process of single-layer PCF-graphene, with initiation of nanofracture at $16.30$\%~of strain and complete nanofracture at $16.55$\% of strain. The small boxes present us an zoomed view of atomistic configuratuion of van der Wall and dynamic bonds from the beginning of the nanofracture until the complete nanofracture of the single-layer PCF-graphene, (d) and (e) at $16.30$\%~of strain. In this theoretical result obtained by reactive CMD simulations showed the representative nanostructural fully atomic model of PCF-graphene single-layer under uniaxial stress load in $y$-direction. Panel (a) in Fig. 8 shows the PCF-graphene single-layer at null strain load. In panel (b) initiation of nanofracture with the breaking of some chemical bonds at $16.30$\%~ of strain, in (c) the single-layer completely nanofractured at $16.55$\%~ of strain, in (d) a zoomed view showed the break of some chemical bonds $C-C$ and (e) a zoomed view in perspective. In the sidebar, the red color indicates high-stress accumulation, while blue color indicates low-stress accumulation in the PCF-graphene single-layer (see Fig.\ref{FIG:PCF-G-Y:03}). However, to verify the mechanical nanofracture pattern in the PCF-graphene single-layer under uniaxial strain load applied in the $y$-direction, we analyzed the stretching evolution of the chemical bonds (see Fig.\ref{FIG:PCF-G-BONDS-STRAIN}). Panels (a)-(d) in Fig. 9 show the beginning of mechanical nanofracture occurs at the bonds $C_{1} - C_{2}$ and $C_{2} - C_{2}$ (see Fig.\ref{FIG:PCF-G-Y:Bonds_lenght} (c) and (d)). These bonds are approximately aligned parallel to the direction of uniaxial mechanical stretching in the $y$-direction (see Fig.\ref{FIG:PCF-G-BONDS-STRAIN}). In computational reactive CMD, the results showed that the chemical bonds between the canbon atoms $C_{1} - C_{1}$ and $C_{2} - C_{2}$, nanofractures do not influence the mechanical failure in PFC-graphene single-layer, (see Fig.\ref{FIG:PCF-G-Y:Bonds_lenght} (a) and (b)). In Fig.\ref{FIG:PCF-G-Y:Diameter}, we showed the pore evolution of PCF-graphene single-layer under uniaxal strain load applied in $y$-direction. We can see that porous large deformations but after
nanofracture of membrane ($3.61$\AA~ up to $4.16$\AA~), return to original nanostructural shape, in approximately $3.59$\AA~ (see Fig.\ref{FIG:PCF-G-Y:Diameter}). In \textbf{supporting information} (attached), we showed the results of snapshots of the stretching process in the $x$-direction and $y$-direction for temperatures of $10$ K, $600$ K, $900$ K and $1200$ K (Figs. \textbf{S1}, \textbf{S3}, \textbf{S5}, \textbf{S7}, \textbf{S9}, \textbf{S11}, \textbf{S13} and \textbf{S15}). The mechanical failure tear is different for the temperatures presented, compered to that presented for $300$ K. In \textbf{supplementary movie S1} (em anexo), one may look at the results for streght (in $x$-direction and $y$-direction) process failure mechanics of PCF-graphene single-layer at temperature $10$ K. We also provide in the supplementary material the atomic coordinates for single-layer PCF-graphene and for the PCF-G-NT nanotubes $(6,0)$ and $(0,7)$ in the extension $.xyz$ (\textbf{see attached supporting information}). The thermodynamic stability of single-layer PCF-graphene and its based nanotubes PCF-G-NT $(6,0)$ and $(0,7)$  is analyzed through the Potential Energy curve ($Kcal/mol/atom$ vs time (fs)). The stability of the non-equilibrium reactive (ReaxFF) classical molecular dynamics simulations was obtained in the statistical ensemble  NVT ({\textit{Nose/Hoover thermostat}}) during a thermalization time $250000$ fs at $300$ K (see \textbf{supporting information:} Fig.\textbf{S17}, Fig.\textbf{S18} and Fig.\textbf{S19} (attached)).

In Fig.\ref{FIG:PCF-G-SS:04}  we present the plots of stress (GPa.\AA) versus strain which indicate the response of mechanical load in $x$-diretion and $y$-direction of the PCF-graphene single-layer based on the reactive (ReaxFF) CMD at $10$, $300$, $600$, $900$ and $1200$ K of temperature.  From the graphical representations of panels (a) and (b) in Fig. 10 we can see two regimes, an elastic and a second plastic regime. The elastic regime (linear region in the graphical), the PCF-graphene under load strain is stretched where it does not break chemical bonds. In the plastic region in PCF-graphene single layer we obtained that the totally nanostructural recovery is no longer possible, resulting from permanent deformations. Nevertheless, the response to mechanical load strain shows  different results for the $x$ and $y$-directions of strain applied and temperature, as well. For strain applied in the $x$ direction at $10$ K, $300$ K and $600$ K, the stress-strain curves of PCF-graphene single-layer are similar to brittle materials and/or nanostructures. While for temperatures of $900$ K and $1200$ K are similar to ductile materials and/or nanostructures. For more information, see the \textbf{supporting information} (attached), where we present snapshots related to the stress-strain curve for a better understanding of the mechanical behavior of PCF-graphene single-layer under uniaxial strain load applied in $x$-direction (see Figs. \textbf{S1} up to \textbf{S8}). Panel (b) in Fig.10 shows a graphical representation of the stress-strain curve for PCF-graphene single-layer in the strain applied in the $y$-direction. For temperature of $300$ K, the results are similar to brittle materials and/or nanostructures, while for $10$ K, $600$ K, $900$ K and $1200$ K, the results are similar to ductile materials and/or nanostructures. In \textbf{supporting information} (attached), where we present snapshots related to the stress-strain curve for a better understanding of the mechanical behavior of PCF-graphene single-layer under uniaxial strain load applied in $y$-direction (see Figs. \textbf{S9} up to \textbf{S16}).  

In Figs.\ref{FIG:PCF-GNT-SS:05} and \ref{FIG:PCF-GNT-SS:06}, we present the plots of stress (GPa.\AA) versus strain which indicate the behavior of the response of uniaxial mechanical load in $z$-diretion of the PCF-graphene-based nanotubes (PCF-G-NTs) $(n,0)$ and $(0,n)$, obtained based on the reactive (ReaxFF) CMD at $10$, $300$, $600$, $900$ and $1200$ K of temperature. As shown, the results are similar to brittle materials and/or nanostructures.  Table 1 shows the structural specifications of all PCF-G-NTs $(0,n)$ and $(n,0)$ studied in this work. Fig.\ref{FIG:PCF-GNTs_snapshots} shows representative snapshots of the nanostructural fully atomic model of PCF-graphene-based nanotubes $(0,7)$ and $(6,0)$ chirality, respectively, under uniaxial stress load applied in $z$-direction. Plots (a)-(c) in Fig.13 show the failure mechanical nanofracture mechanics of PCF-G-NT $(0,7)$ where (a), (b) and (c) are at null strain load, fully stressed at $20.27$\%~ and  completely nanofractured at $20.74$\% of strain load, respectively. The small boxes (e) and (f) in Fig.13 show the break of some chemical bonds $C-C$ and PCF-G-NT $(0,7)$ completely fractured, respectively. In plots (d) and (g) a perspective view of the PCF-G-NT $(0,7)$ presented. In addition, plots (h) - (n) present the failure mechanical nanofracture mechanics of PCF-G-NT $(6,0)$ so that (h), (i) and (j) are at null strain load, fully stressed at $34.41$\%~ and completely nanofractured at $20.74$\% of strain load, respectively. Furthermore, the small boxes (l) and (m) in Fig.13 show the break of some chemical bonds $C-C$ and PCF-G-NT $(0,7)$ completely fractured, respectively. Finaly, the plots (k) and (n) present a perspective view of the PCF-G-NT $(6,0)$.

Fig.\ref{FIG:PCF-G-NT_PR}, shows the graphical Poisson’s coefficients representation of PCF-G-NTs $(6,0)$ and $(0,7)$ as a function of the uniaxial strain applied in $z$-direction. After subjecting PCF-G-NT $(6,0)$ to a $14$\%~ of load mechanical strain, a noticeable decline in the Poisson's coefficient is observed, eventually reaching a remarkable value of $0.20$. Conversely, PCF-G-NT $(0,7)$ Poisson's coefficient even under the same strain load conditions is $0.48$.

Using the Poisson’s coefficients, we can obtain an elastic property of the PCF-graphene-based nanotubes which studied in this work. The evaluations were done by transversal deformation of the PCF-G-NTs, with homogeneous and isotropic fully atomistic carbon atoms configurations. The results showed that the values for the Poisson's ratio are different between PCF-G-NT $(6,0)$ and $(0,7)$. This finding is consistent with the fact that PCF-G-NT $(6,0)$ and $(0,7)$ show different nanostructural morphologies.

\section{Conclusions and remarks}
\label{conclusion}

The present study deals the mechanical properties and nanofracture patterns of PCF-graphene single-layer  and PCF-graphene-based nanotubes (PCF-G-NT $(n,0)$ and $(0,n)$), performed by reactive (ReaxFF) Classical Molecular Dynamics Simulations method. The stress-strain graphical representation behavior of the PCF-graphene was observed to follow two regimes: $(i)$ one exhibiting linear elasticity and $(ii)$ a plastic ones (involving carbon atom re-hybridization and bond breaking). The PCF-graphene single-layer for strain applied in the $x$-direction at $10$ K, $300$ K and $600$ K, are similar to brittle materials and/or
nanostructure and for $900$ K and $1200$ K are similar to ductile materials and/or nanostructures. The PCF-graphene single-layer whith strain applied in the $y$-direction ate temperature of the $300$ K, are similar to brittle materials and/or nanostructures, while for $10$ K, $600$ K, $900$ K and $1200$ K, are similar to ductile materials and/or nanostructures. The Young's Modulus for  PCF-graphene single-layer for uniaxial strain in the $x$-direction ranges from $5651.7 - 4328.6$ GPa.\AA~ and in the $y$-direction $2408.5 - 1934.4$ GPa\AA~, considering the thickness of $3.35$\AA~. The Young's Modulus for the PCF-graphene single-layer is $40.99$\%~ larger than the graphene single-laye. The Young's modulus of the  PCF-G-NTs ($(0, n)$ and $(n, 0)$) are range $1850.5 - 2603.3$ GPa\AA~ and, $386.25 - 1280.7$ GPa\AA, respectively. The Poisson’s coefficients value representation of PCF-G-NTs $(6,0)$ and $(0,7)$ as a function of the uniaxial strain applied in $z$-direction ( $30$\%~ of load mechanical strain) are $0.68$ and  $0.48$, respectively.

\section{Acknowledgements}

This work was supported in part by the Brazilian Agencies CAPES, CNPq, FAPESP and FAPEPI.  J.M.S acknowledges CENAPAD-SP (Centro Nacional de Alto Desenpenho em São Paulo - Universidade Estadual de Campinas - UNICAMP) provided computational support (proj842). R. Kalami, S.A. Ketabi and J.M. De Sousa would like to thank School of Physics, Damghan University, Condensed Matter Physics department.

\newpage
\bibliographystyle{elsarticle-num}
\bibliography{bibliografia.bib}

\newpage
\begin{table}[htb!]
    \centering
        \caption{Structural parameters of the nanostructural model of PCF-graphene-based nanotubes (PCF-G-NTs) simulated by the reactive Classical Molecular Dynamics Method. The chirality, number of carbon atoms, diameter (\AA) and length (\AA).}
    \begin{tabular}{|c|c|c|c|c|}
    \hline
         Chirality & $n$ & Number of Atoms & Diameter (\r{A}) & Length (\r{A}) \\ \hline
        \multirow{6}{*}{ $(n,0)$}& 4     & 384		& 6.26		& 32.92 \\ \cline{2-5}
        & 5 	& 480		& 7.82		& 32.92 \\ \cline{2-5}
        & 6    & 576		& 9.39		& 32.92 \\  \cline{2-5}
        & 7 	& 672		& 10.95		& 32.92 \\ \cline{2-5} 
        & 8 	& 768		& 12.52		& 32.92 \\ \cline{2-5}
        & 9 	& 864		& 14.08		& 32.92  \\ \cline{2-5}
        & 10 	& 960		& 15.64		& 32.92 \\ \cline{2-5}\hline
        \multirow{6}{*}{ $(0,n)$} &  4	&  448		& 6.99		& 34.40 \\ \cline{2-5}
        & 5	& 560		& 8.73		& 34.40 \\ \cline{2-5}
        & 6	& 672		& 10.48		& 34.40 \\ \cline{2-5}
        & 7	& 784		& 12.23		& 34.40 \\ \cline{2-5}
        & 8	& 896		& 13.97		& 34.40 \\ \cline{2-5}
        & 9	& 1008		& 15.72		& 34.40 \\ \cline{2-5}
        & 10	& 1120		& 17.47		& 34.40 \\ \hline
    \end{tabular}
    \label{tab:PCF-NTs-tab}
\end{table}

\begin{table}[htb!]
    \centering
        \caption{Young's Modulus (GPa.\AA), Ultimate Tensile Strength (GPa.\AA), critical strain $\epsilon_{C}$(\%), for PCF-graphene-based single-layer and nanotubes (PCF-G-NTs)}
    \begin{tabular}{|c|c|c|c|c|}
    \hline
    \multicolumn{5}{|c|}{PCF-graphene single-layer} \\ \hline
     strain direction &  temperature (K) & $Y_{Mod}$ (GPa.\AA) & UTS (GPa.\AA) & $\sigma_C$ (\%)  \\ \hline
      \multirow{5}{*}{$x$}
        & 10   & 5651.7 $\pm$ 42.51 & 917.68 & 11.12 \\ \cline{2-5}
        & 300  & 4839.7 $\pm$ 34.99 & 765.89 & 10.43 \\ \cline{2-5}
        & 600  & 4570.3 $\pm$ 42.32 & 654.76 & 10.33 \\ \cline{2-5}
        & 900  & 4363.4 $\pm$ 50.10 & 526.58 & 18.71 \\ \cline{2-5}
        & 1200 & 4328.6 $\pm$ 53.76 & 448.74 & 14.87 \\ \hline
       \multirow{5}{*}{$y$} 
        & 10   & 2408.5 $\pm$ 8.84 & 690.50 & 25.13 \\ \cline{2-5}
        & 300  & 2231.5 $\pm$ 15.77 & 551.41 & 16.51 \\ \cline{2-5}
        & 600  & 2212.5 $\pm$ 18.30 & 489.03 & 33.37 \\ \cline{2-5}
        & 900  & 2036.4 $\pm$ 17.81 & 427.37 & 29.83 \\ \cline{2-5}
        & 1200 & 1934.4 $\pm$ 18.12 &394.17 & 28.36 \\ \hline
        \multicolumn{5}{|c|}{PCF-graphene-based nanotubes (PCF-G-NTs) ($300$ K)} \\ \hline
        chirality &   $(n,0),(0,n)$ & $Y_{Mod}$ (GPa.\AA) & UTS (GPa.\AA) & $\sigma_C$ (\%)  \\ \hline
        \multirow{6}{*}{$(n,0)$}
        & (4,0)   & 386.25 $\pm$ 17.55 & 123.78 & 22.22   \\ \cline{2-5}
        & (5,0)   & 485.47 $\pm$ 20.95 & 123.98  & 39.27   \\ \cline{2-5}
        & (6,0)   & 543.54 $\pm$ 31.64 & 223.94 & 28.10  \\ \cline{2-5}
        & (7,0)   & 688.34 $\pm$ 11.59 & 159.78 & 25.76   \\ \cline{2-5}
        & (8,0)   & 1246.7 $\pm$ 22.57 & 266.00 & 21.75   \\ \cline{2-5}
        & (9,0)   & 1093.1 $\pm$ 11.59 & 247.35 & 28.98   \\ \cline{2-5}
        & (10,0)   & 1280.7 $\pm$  10.13 & 316.83 & 22.70   \\ \hline \hline
        \multirow{6}{*}{$(0,n)$}
        & (0,4)  & 1850.5 $\pm$ 58.14 & 382.27 & 12.23   \\ \cline{2-5}
        & (0,5)  & 1982.0 $\pm$ 52.77 & 451.72 & 25.87   \\ \cline{2-5}
        & (0,6)  & 1995.4 $\pm$ 26.41 & 379.48 & 18.12   \\ \cline{2-5}
        & (0,7) & 2575.3 $\pm$  27.95 & 543.54 & 20.34   \\ \cline{2-5} 
        & (0,8) & 2514.4 $\pm$ 37.82 & 593.18 & 21.13   \\ \cline{2-5}
        & (0,9) & 2731.6 $\pm$  32.31 & 592.10 & 21.16   \\ \cline{2-5}
        & (0,10) & 2603.3 $\pm$  33.65 & 588.15 & 24.07   \\ \cline{2-5}\hline
    \end{tabular}
    \label{tab:elastic-val}
\end{table}

\begin{figure}[ht!]
\begin{center}
\includegraphics[angle=0,scale=0.22]{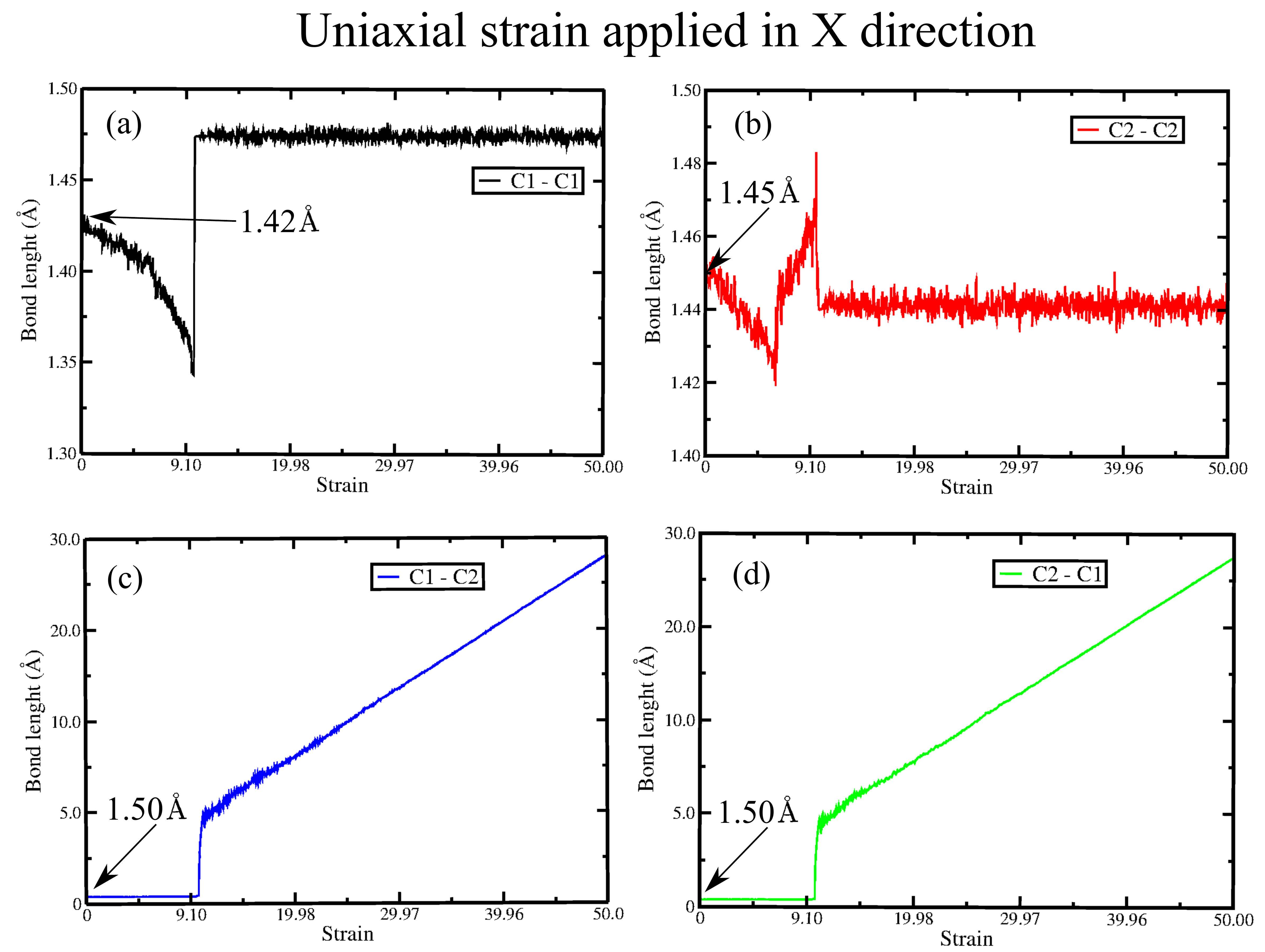}
\caption{\footnotesize{ Bond length evolution with strain for PCF-graphene single-layer whith uniaxial strain applied in $X$ direction at at  temperature $10$ K. In (a) chemical bond strain evolution for $C_{1} - C_{1}$, in (b) for $C_{2} - C_{2}$, (c) $C_{1} - C_{2}$ and (d) $C_{2} - C_{1}$.  In Figure \ref{FIG:PCF-G-BONDS-STRAIN} is indicated the labeling of each $C_{i}$ carbon atom.}}
\label{FIG:PCF-G-X:Bonds_lenght}
\end{center}
\end{figure}

\begin{figure}[ht!]
\begin{center}
\includegraphics[angle=0,scale=0.18]{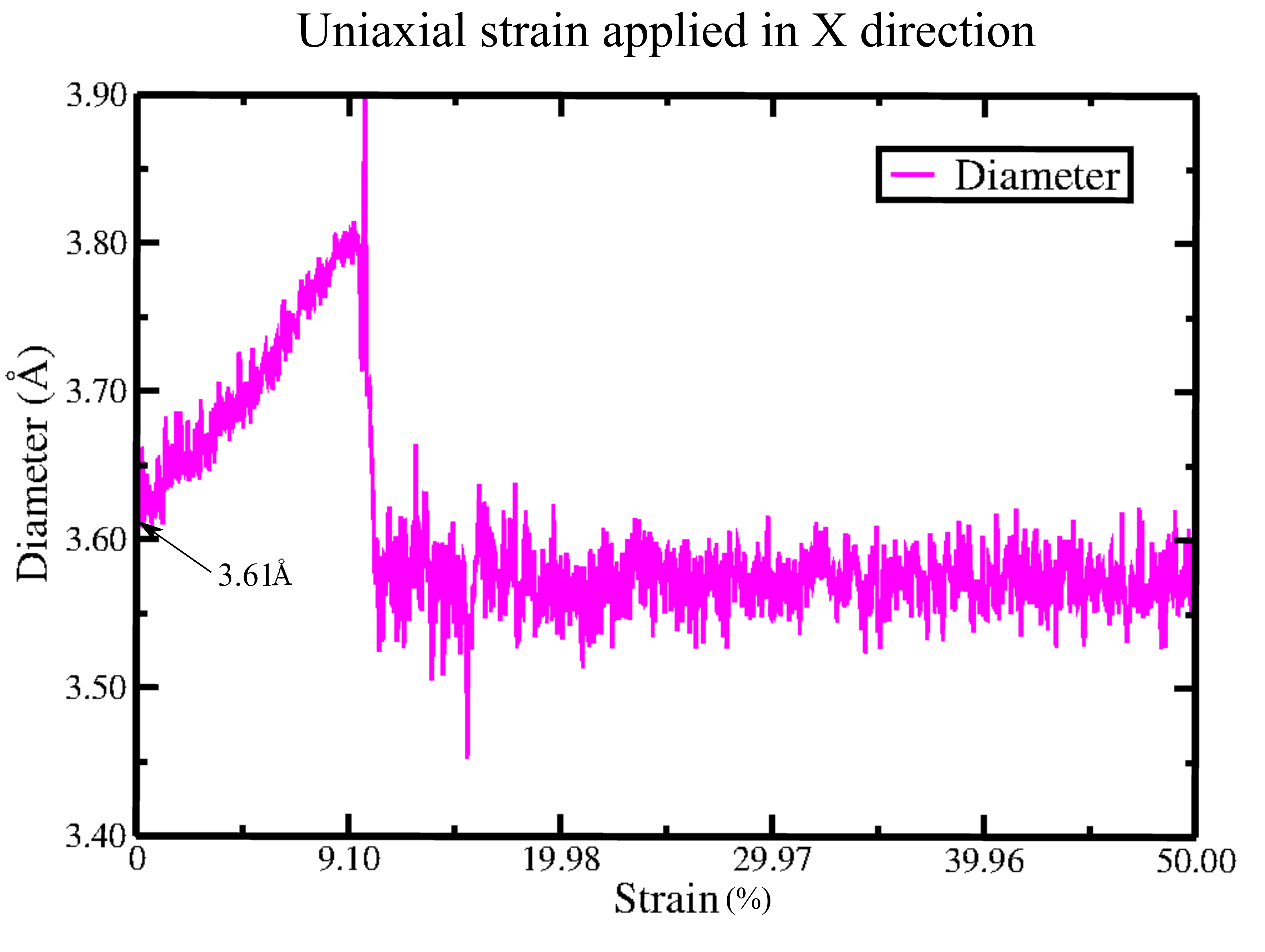}
\caption{\footnotesize{ Diameter length evolution with strain for PCF-graphene single-layer whith uniaxial strain applied in $X$ direction at $10K$ temperature. See Figure \ref{FIG:PCF-G-BONDS-STRAIN} is indicated the diameter represented by the circle in red color. }}
\label{FIG:PCF-G-X:Diameter}
\end{center}
\end{figure}

\begin{figure}[ht!]
\begin{center}
\includegraphics[angle=0,scale=0.27]{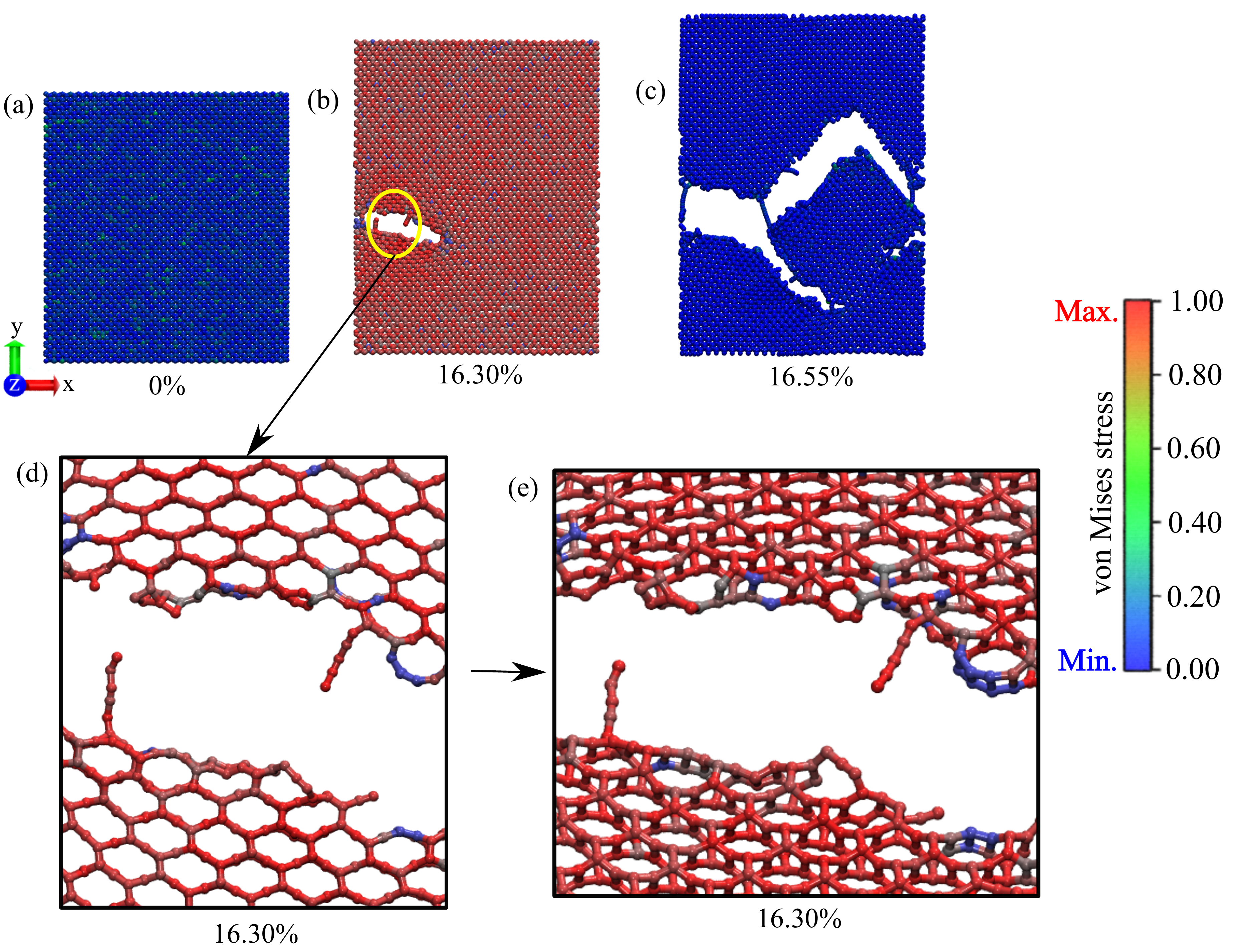}
\caption{\footnotesize{Representative nanostructural fully atomic model of PCF-graphene single-layer under uniaxial stress load in $Y$ direction. In (a) at $0$\% of strain, in (b) initiation of nanofracture with the breaking of some chemical bonds at $16.30$\% of strain, in (c) the single-layer completely nanofractured at $16.55$\% of strain, in (d) a zoomed view showed the break of some chemical bonds $C - C $ and (e) a zoomed view in perspective. In the sidebar located on the right side, the red color indicates high-stress accumulation, while blue color indicates low-stress accumulation in the PCF-graphene single-layer.}}
\label{FIG:PCF-G-Y:03}
\end{center}
\end{figure}

\begin{figure}[ht!]
\begin{center}
\includegraphics[angle=0,scale=0.22]{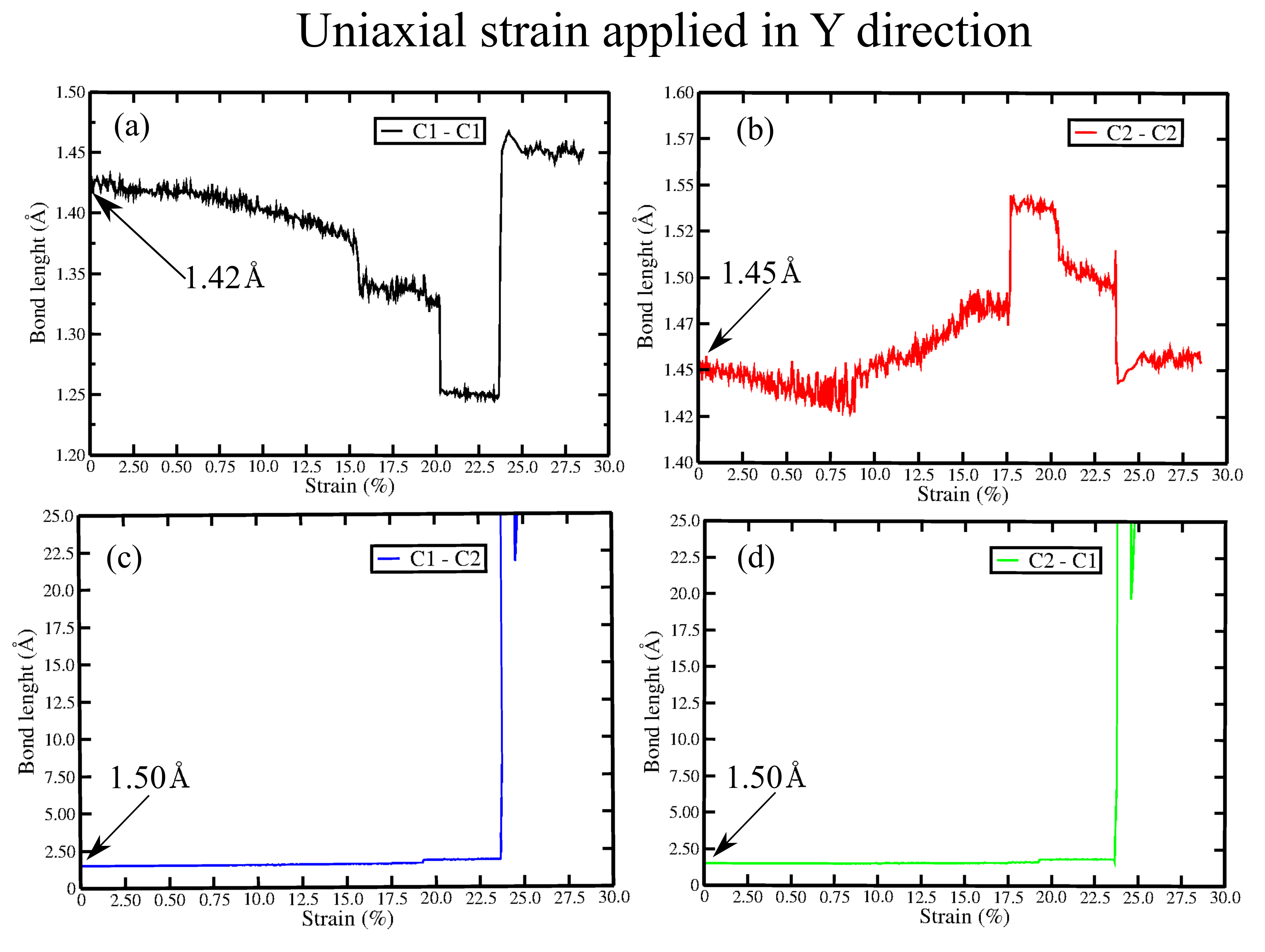}
\caption{\footnotesize{ Bond length evolution with strain for PCF-graphene single-layer whith uniaxial strain applied in $Y$ direction at  temperature $10$ K. In (a) chemical bond strain evolution for $C_{1} - C_{1}$, in (b) for $C_{2} - C_{2}$, (c) $C_{1} - C_{2}$ and (d) $C_{2} - C_{1}$.  In Figure \ref{FIG:PCF-G-BONDS-STRAIN} is indicated the labeling of each $C_{i}$ carbon atom.}}
\label{FIG:PCF-G-Y:Bonds_lenght}
\end{center}
\end{figure}

\begin{figure}[ht!]
\begin{center}
\includegraphics[angle=0,scale=0.18]{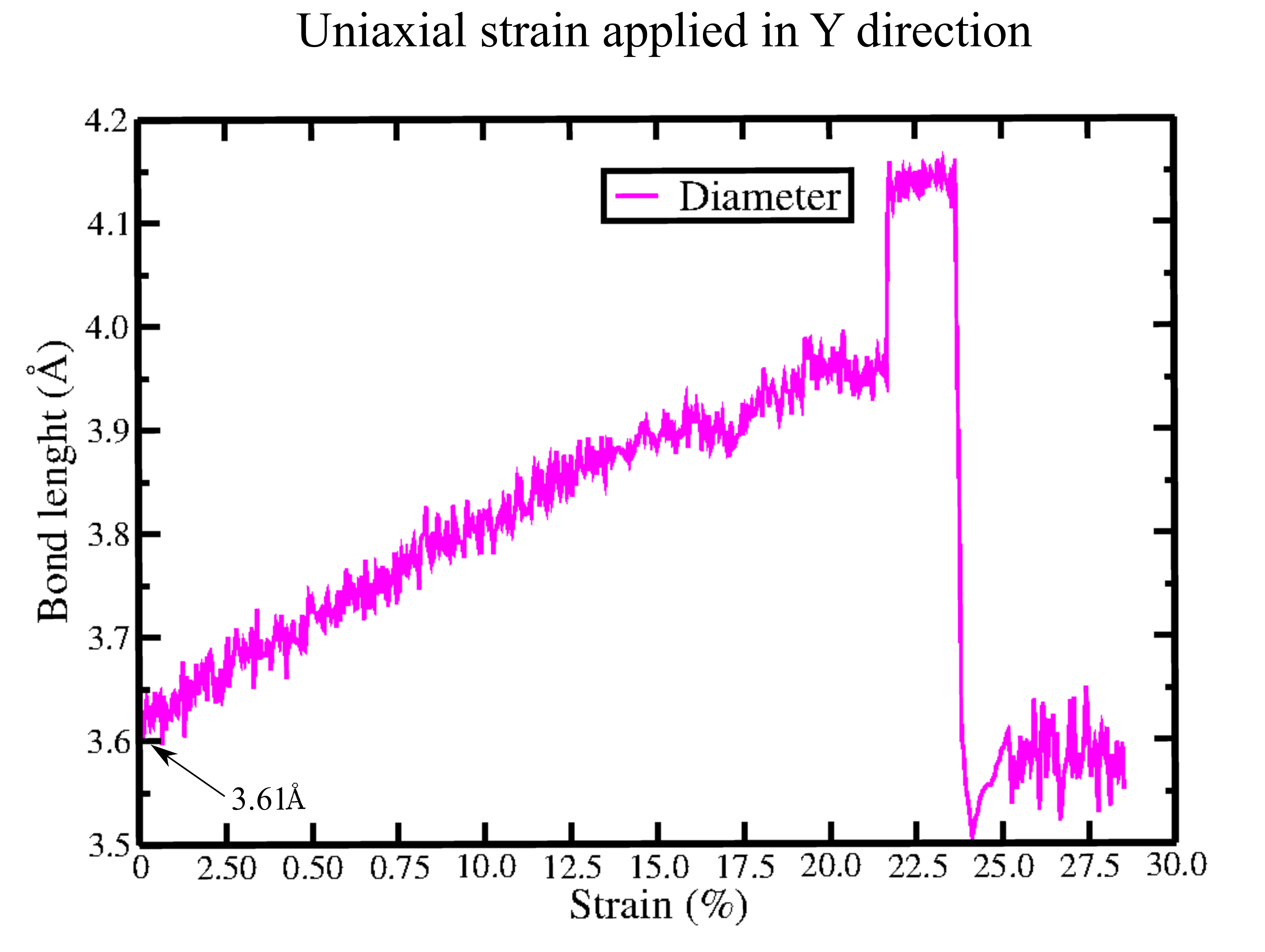}
\caption{\footnotesize{ Diameter length evolution with strain for PCF-graphene single-layer whith uniaxial strain applied in $Y$ direction at $10K$ temperature. See Figure \ref{FIG:PCF-G-BONDS-STRAIN} is indicated the diameter represented by the circle in red color. }}
\label{FIG:PCF-G-Y:Diameter}
\end{center}
\end{figure}

\begin{figure}[ht!]
\begin{center}
\includegraphics[angle=0,scale=0.26]{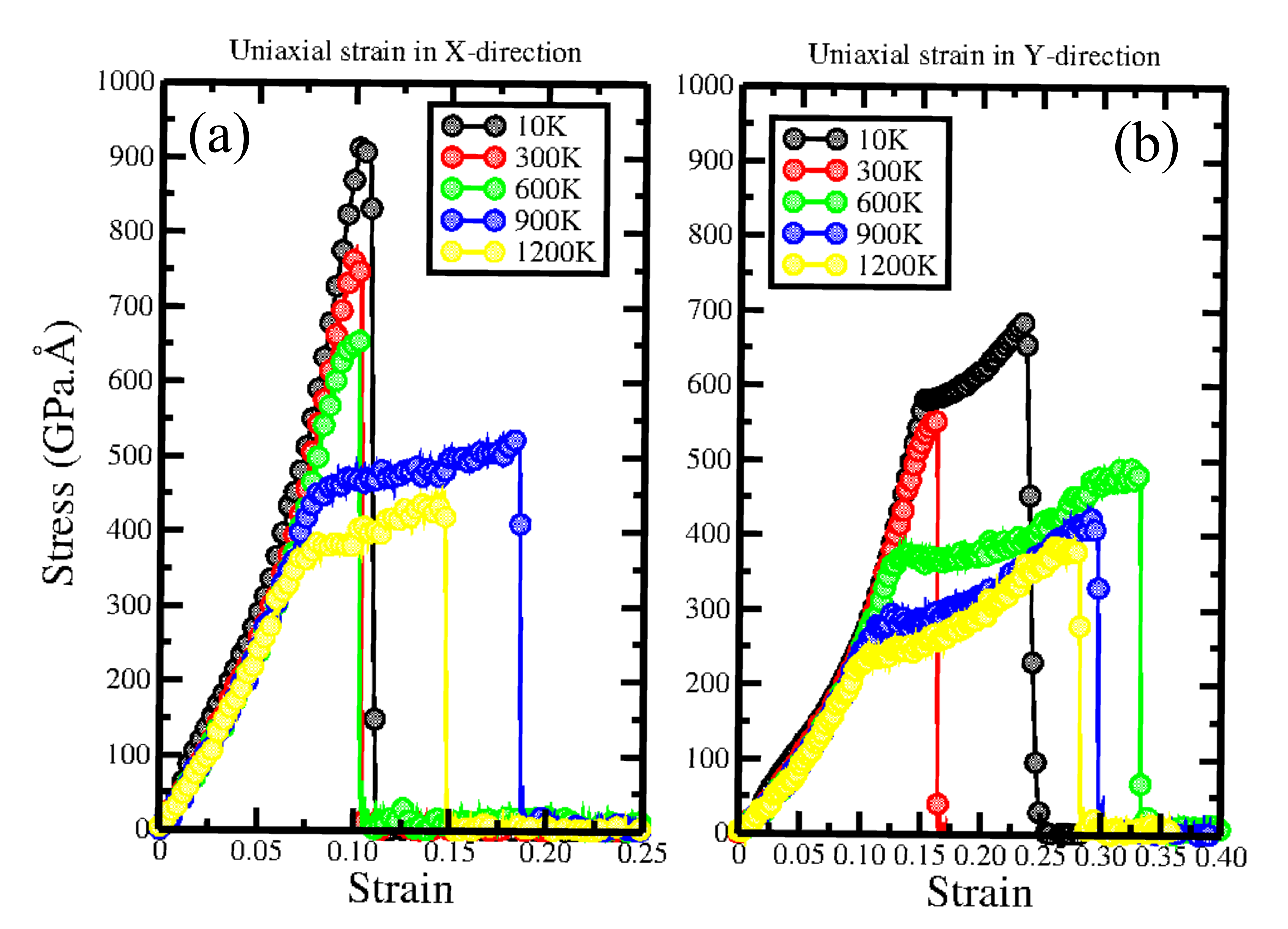}
\caption{\footnotesize{ PCF-graphene single-layer stress-strain curve predicted by Classical Molecular Dynamics Simulations Method calculations. In (a) we have the results of the stress-strain curves for uniaxial stress applied in the $X$ direction for temperatures of $10$K, $300$K, $600$K, $900$K and $1200$K. In (b) we have the results of the stress-strain curves for uniaxial stress applied in the $Y$ direction for the same temperatures.}}
\label{FIG:PCF-G-SS:04}
\end{center}
\end{figure}

\begin{figure}[ht!]
\begin{center}
\includegraphics[angle=0,scale=0.35]{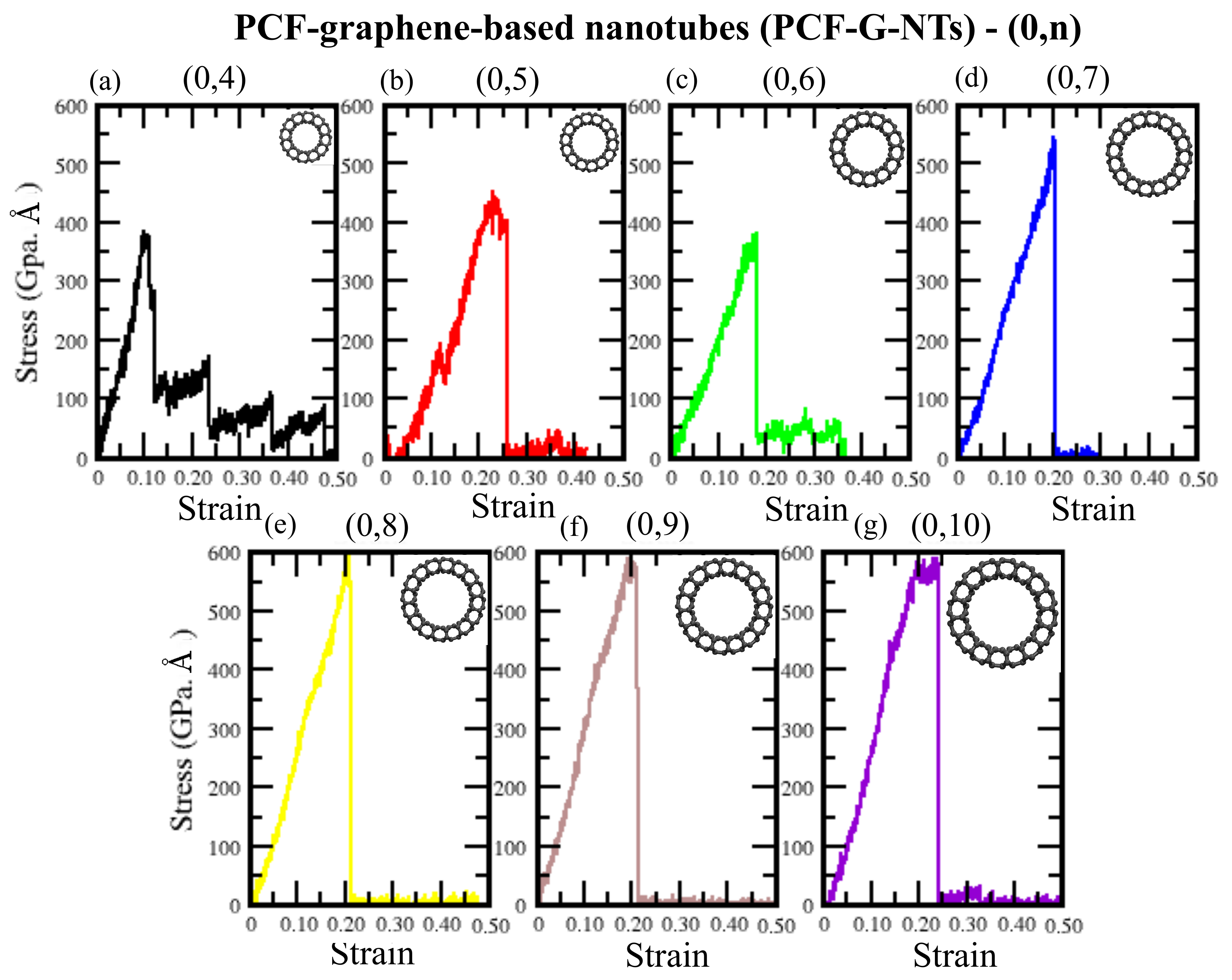}
\caption{\footnotesize{ PCF-graphene-based nanotubes (PCF-G-NTs) stress-strain curve predicted by Classical Molecular Dynamics Simulations Method calculations.  We have the results of the stress-strain curves for uniaxial stress applied in the $Z$ direction at room temperature. (a) up to (g) PCF-G-NTs $(0,4)$ up to $(0,10)$ chirality, respectively.}}
\label{FIG:PCF-GNT-SS:05}
\end{center}
\end{figure}

\begin{figure}[ht!]
\begin{center}
\includegraphics[angle=0,scale=0.35]{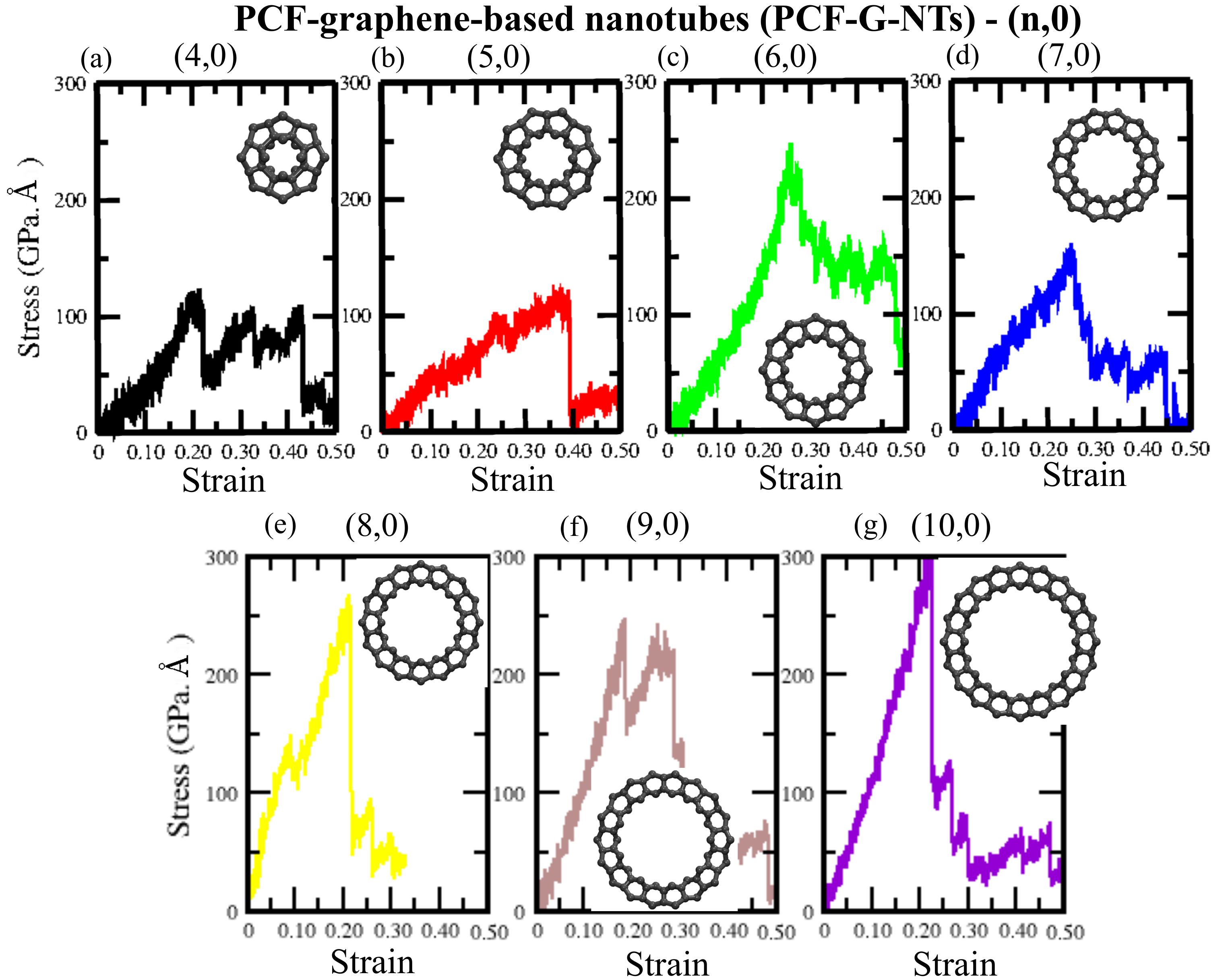}
\caption{\footnotesize{ PCF-graphene-based nanotubes (PCF-G-NTs) stress-strain curve predicted by Classical Molecular Dynamics Simulations Method calculations.  We have the results of the stress-strain curves for uniaxial stress applied in the $Z$ direction at room temperature. (a) up to (g) PCF-G-NTs $(4,0)$ up to $(10,0)$ chirality, respectively.}}
\label{FIG:PCF-GNT-SS:06}
\end{center}
\end{figure}

\begin{figure}[ht!]
\begin{center}
\includegraphics[angle=0,scale=0.18]{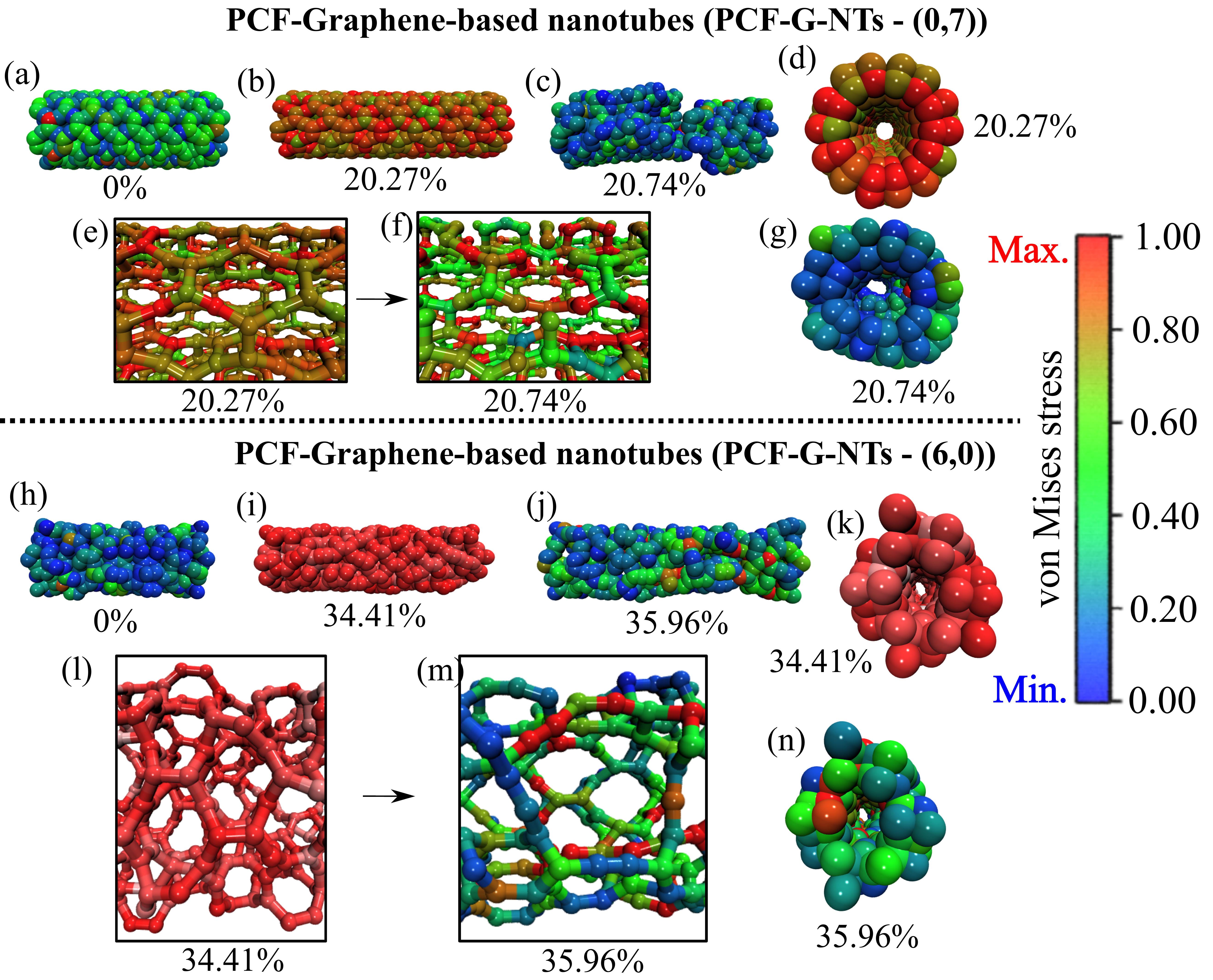}
\caption{\footnotesize{ Representative nanostructural fully atomic model of PCF-graphene-based nanotubes $(0,7)$ and $(6,0)$ chirality, respectively, under uniaxial stress load applied in
$Z$ direction. In (a) PCF-G-NTs $(0,7)$ at $0$\% of strain, in (b) highly tensioned at  $20.27$\% of strain, in (c) the PCF-G-NTs $(0,7)$ completely nanofractured at $20.74$\% of strain, in (d) and (g) a perspective view of PCF-G-NTs $(0,7)$ at $20.75$\% and $20.74$\% of strain, respectively. In (e) the PCF-G-NTs $(0,7)$ with high tensile load showing chemical bonds before mechanical nanofracture, in (f) showed the nanofracture pattern  of  PCF-G-NTs $(0,7)$. In (h) PCF-G-NTs $(6,0)$ at $0$\% of strain, (i) highly tensioned at  $34.41$\% of strain, in (j)  the PCF-G-NTs $(6,0)$ completely nanofractured at $35.96$\% of strain, in (k) and (n) a perspective view of PCF-G-NTs $(6,0)$ at $34.41$\% and $35.96$\% of strain, in (l) the PCF-G-NTs $(6,0)$ with high tensile load showing chemical bonds before mechanical nanofracture at $34.35$\% of atrain and, (m) showed the nanofracture pattern of PCF-G-NTs $(6,0)$.}}
\label{FIG:PCF-GNTs_snapshots}
\end{center}
\end{figure}

\begin{figure}[ht!]
\begin{center}
\includegraphics[angle=0,scale=0.30]{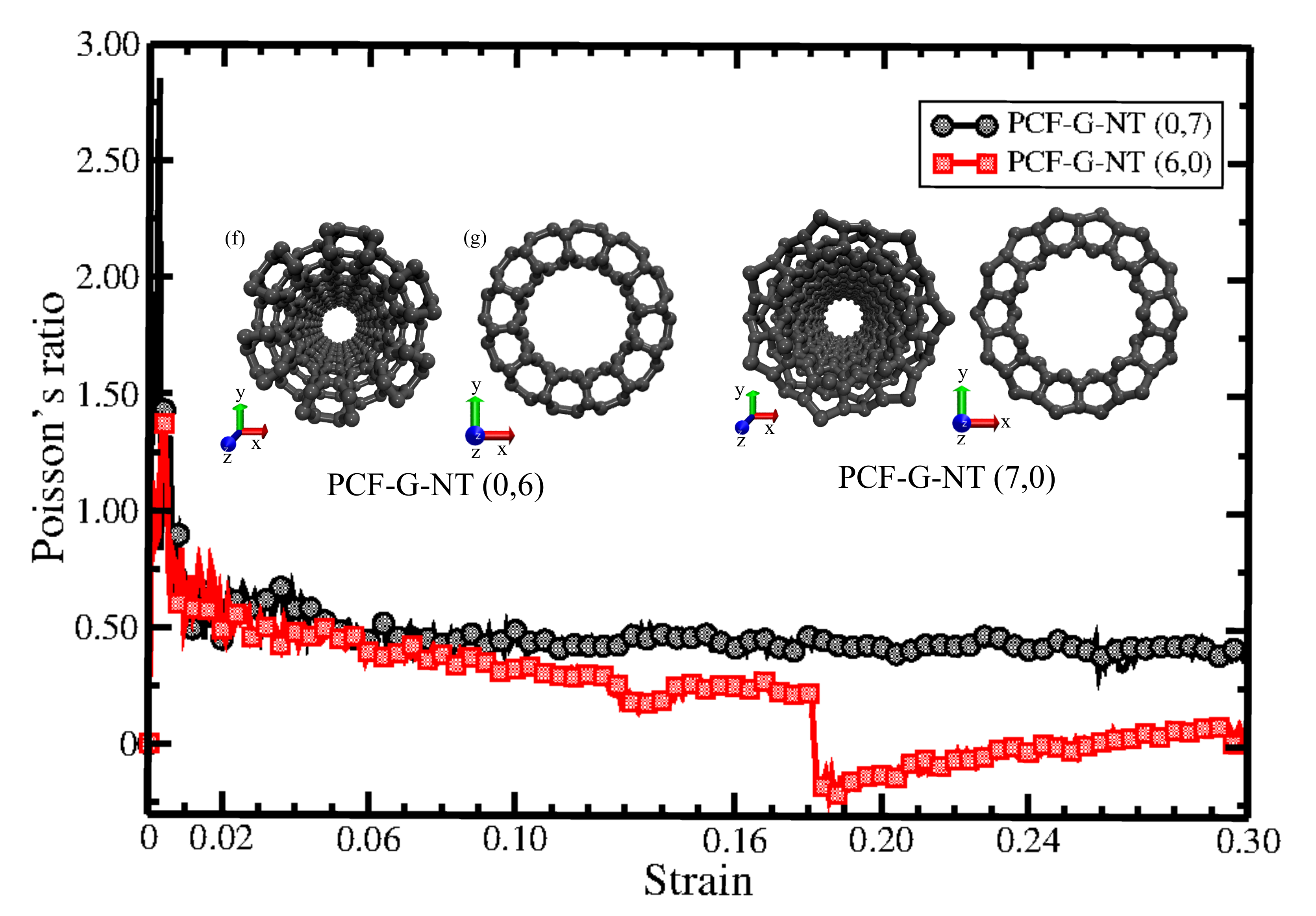}
\caption{\footnotesize{Graphical representation of the Posisso'n ratio  versus strain obtained by the reactive (ReaxFF) classical molecular dynamics simulations method at room temperature for the PCF-G-NTs $(6,0)$ (red color) and $(0,7)$ (black color).}}
\label{FIG:PCF-G-NT_PR}
\end{center}
\end{figure}

\end{document}